\documentclass[
  journal=pasa,
  manuscript=article-type,
  year=2025,
  volume=1,
]{cup-journal}

\usepackage{amsmath}
\usepackage[nopatch]{microtype}
\usepackage{bbm}
\usepackage{booktabs}
\usepackage{amssymb}
\usepackage{commath}

\usepackage{tikz,xcolor,hyperref}

\newcommand{\direct}{\textsc{DIReCT}}
\newcommand{\gendirect}{\textsc{GenDIReCT}}

\newcommand{\ehtim}{\textsc{eht-imaging}}

\newcommand{\clean}{\textsc{CLEAN}}

\newcommand{\smili}{\textsc{SMILI}}

\defcitealias{Chael_2018_ehtim}{C18}
\defcitealias{Roelofs2023_ngeht-challenge}{R23}

\definecolor{lime}{HTML}{A6CE39}
\DeclareRobustCommand{\orcidicon}{%
    \begin{tikzpicture}
    \draw[lime, fill=lime] (0,0) 
    circle [radius=0.16] 
    node[white] {{\fontfamily{qag}\selectfont \tiny ID}};
    \draw[white, fill=white] (-0.0625,0.095) 
    circle [radius=0.007];
    \end{tikzpicture}
    \hspace{-2mm}
}

\newcommand{\orcidSamuel}{\href{https://orcid.org/0000-0001-9372-4611}{\orcidicon}}
\newcommand{\orcidNT}{\href{https://orcid.org/0000-0003-1602-7868}{\orcidicon}}

\title[Diffusion VLBI Imaging]{Very-Long Baseline Interferometry Imaging with Closure Invariants using Conditional Image Diffusion}

\author{Samuel Lai\orcidSamuel}
\affiliation{Space \& Astronomy, Commonwealth Scientific and Industrial Research Organisation (CSIRO), P. O. Box 1130, Bentley, WA 6102, Australia}
\email[Samuel Lai]{samuel.lai@csiro.au}

\author{Nithyanandan Thyagarajan\orcidNT}
\affiliation{Space \& Astronomy, Commonwealth Scientific and Industrial Research Organisation (CSIRO), P. O. Box 1130, Bentley, WA 6102, Australia}

\author{O. Ivy Wong}
\affiliation{Space \& Astronomy, Commonwealth Scientific and Industrial Research Organisation (CSIRO), P. O. Box 1130, Bentley, WA 6102, Australia}
\alsoaffiliation{International Centre for Radio Astronomy Research, The University of Western Australia, Crawley, WA 6009, Australia}

\author{Foivos Diakogiannis}
\affiliation{Data 61, Commonwealth Scientific and Industrial Research Organisation (CSIRO), Kensington, WA 6151, Australia}


\begin{document}

\begin{abstract}
Image reconstruction in very-long baseline interferometry operates under severely sparse aperture coverage with calibration challenges from both the participating instruments and propagation medium, which introduce the risk of biases and artefacts. Interferometric closure invariants offers calibration-independent information on the true source morphology, but the inverse transformation from closure invariants to the source intensity distribution is an ill-posed problem. In this work, we present a generative deep learning approach to tackle the inverse problem of directly reconstructing images from their observed closure invariants. Trained in a supervised manner with simple shapes and the CIFAR-10 dataset, the resulting trained model achieves reduced chi-square data adherence scores of $\chi^2_{\rm CI} \lesssim 1$ and maximum normalised cross-correlation image fidelity scores of $\rho_{\rm NX} > 0.9$ on tests of both trained and untrained morphologies, where $\rho_{\rm NX}=1$ denotes a perfect reconstruction. We also adapt our model for the Next Generation Event Horizon Telescope total intensity analysis challenge. Our results on quantitative metrics are competitive to other state-of-the-art image reconstruction algorithms. As an algorithm that does not require finely hand-tuned hyperparameters, this method offers a relatively simple and reproducible calibration-independent imaging solution for very-long baseline interferometry, which ultimately enhances the reliability of sparse VLBI imaging results. 
\end{abstract}

\section{Introduction} \label{sec:Introduction}

Very long baseline radio interferometry (VLBI) is a technique used to measure spatial correlations from distant signals, leveraging baselines on continental or planetary scales to probe extremely small angular scales ($\sim 10\,\mu$as). The measured correlations between receiver elements in the array, known as visibilities, can be used to infer the source intensity distribution \citep{TMS}. Naturally, data produced by a limited number of receiver elements separated on such large spatial scales implies severely sparse coverage on the aperture plane, necessitating accurate signal calibration from the heterogeneous array elements for restoration of signal coherence and accurate image recovery. 

Calibration of a VLBI measurement dataset is often a meticulously fine-tuned iterative process that involves converging on the calibration corrections required to produce model visibilities consistent with the observation from iteratively refined model images. Different assumptions during the calibration process can have significant effects on fine details of the image reconstruction \citep[e.g.][]{EHT_2019_Imaging}. For instance, on the observations of M87 \citep{EHT_2019_Data} produced by the Event Horizon Telescope Collaboration \citep[EHTC;][]{Doeleman_2009}, \citet{Carilli_2022} showed that diversity in structures can be produced by a hybrid mapping reconstruction algorithm by adopting different initial models. Other independent analyses of the EHTC data on M87 \citep[e.g.][]{Arras_2022, Broderick_2022, Carilli_2022, Lockhart_2022, Muller_2024_closureTraces, Feng_2024} generally produce consistent ring-like morphologies, but finer characteristics (i.e. on the ring thickness, and surface brightness dynamic range) can vary substantially between various methodologies or prior assumptions. This underscores the importance of an accurate calibration, wherein minute errors can sometimes lead to divergent results.

Alternatively, much of the complexities of the calibration process can be bypassed by considering specific combinations of interferometric measurements. These combinations, known as ``closure quantities'', are traditionally split into closure phases \citep{Jennison_1958,Thyagarajan_2022_CPhase} and closure amplitudes \citep{Twiss_1960}. As closure quantities are immune to  multiplicative station-based noise properties, they can serve as calibration-independent true observables, carrying robust information on source properties limited only by additive thermal noise and non-station-based errors \citep{Blackburn_2020, Lockhart_2022}. Moreover, closure quantities carry robust information on a source's intrinsic polarisation \citep{Broderick_2020, Samuel_2022}. Therefore, the inverse transformation from closure quantities to source morphology is of considerable interest. Regularised maximum-likelihood methods \citep[RML; e.g.][]{Ikeda_2016, Akiyama_2017_polarimetric, Akiyama_2017_imaging, Chael_2018_ehtim, Blackburn_2020} have shown that minimising objective functions containing closure data terms produces calibration-independent images. Similarly, closure quantities can be used as a constraint in compressive sensing techniques, which encode the image with a sparse set of multi-scale basis functions \citep[e.g.][]{Mertens_2015, Mueller_2022_doghit, Muller_2024_closureTraces}.

Recently, noteworthy advances in machine learning have attracted attention for their integration into problems relevant to radio astronomy. A large variety of machine learning techniques have been applied successfully for interferometric image reconstruction, from classification networks \citep[e.g.][]{Thyagarajan_2024_Lucas}, normalising flow \citep[e.g.][]{Sun_2020, Sun_2022_adeep, Feng_2024}, and vision transformers \citep[e.g.][]{Lai_2025_DIReCT}. Our work builds off of the proof-of-concept study of \citet{Lai_2025_DIReCT}, which showed that generalised closure quantities in the co-polar formalism described in \citet{Thyagarajan_2022_CI} can be leveraged directly for image reconstruction by exploiting a transformer's attention mechanism to extract meaningful features from global interrelationships within the closure dataset. In this work, we present a generative deep learning solution for image reconstruction in the VLBI regime, which we call \gendirect, consisting of a conditional diffusion model with an unsupervised convolutional neural network (CNN). We validate \gendirect\ with a variety of tests on trained and untrained morphologies, and various levels of noise. Furthermore, we apply the method on the synthetic datasets of the Next Generation Event Horizon Telescope (ngEHT) total intensity analysis challenge \citep{Roelofs2023_ngeht-challenge} and quantitatively compare our performance to other submissions which use state-of-the-art reconstruction algorithms. 

The content of this paper is organised as follows: in sections \ref{sec:vlbi-data} and \ref{sec:CI-formalism}, we introduce the co-polar data products of VLBI radio observations and the closure invariant formalism, respectively. In section \ref{sec:imaging}, we describe established image reconstruction methods and present our generative machine learning approach, which estimates the conditional image distribution given closure invariants as input. We also describe the training strategy for the deep generative model. In sections \ref{sec:results} and \ref{sec:discussion}, we present and validate the results of the proposed image reconstruction pipeline, applying our method to the ngEHT total intensity analysis challenge datasets and comparing with other reconstruction methods. We present a summary of this work and conclusion in section \ref{sec:conclusion}. 

\section{Primer on VLBI Data Products} \label{sec:vlbi-data}
Radio interferometers measure the complex visibility function, $\mathcal{V}(\mathbf{u})$, representing the spatial coherence of received signals. The van Cittert-Zernike theorem, that relates the visibilities to the sky intensity distribution, under some approximations can be simplified to a two-dimensional Fourier transform \citep{TMS}. 
If we consider that the observed visibilities, $\mathcal{V}'_{ab}$, measured between a pair of stations, $(a,b)$, can be corrupted by complex station-based factors, $g$, representing both amplitude and phase distortions, then we can represent the relationship between true and measured visibilities as $\mathcal{V}'_{ab} = g_a \mathcal{V}_{ab} g_b^* + \epsilon_{ab}$. Here, $\epsilon_{ab}$ is an additive thermal noise term represented by a zero-mean Gaussian. 

Special interferometric ``closure quantities'' can be constructed from combinations of Fourier components, which eliminate the complex station-based multiplicative terms \citep{TMS}. Traditionally, closure phases \citep{Jennison_1958,Thyagarajan_2022_CPhase} and amplitudes \citep{Twiss_1960} are constructed from closed triangular and quadrilateral loops of array elements, respectively. For $N_s$ stations, the total number of $(N_s-1)(N_s-2)/2$ closure phases and $N_s(N_s-3)/2$ closure amplitudes is always fewer than the total number of real-valued visibilities by a factor of 
$1-\frac{1}{N_s}-\frac{1}{N_s-1}$,
resulting in a loss of information, notably of the absolute position and total flux density of the source. However, closure quantities are robust against multiplicative station-based gains, providing calibration-independent information on the source morphology. Furthermore, closure quantities also carry information on a source's intrinsic polarisation characteristics \citep[e.g.][]{Broderick_2020, Samuel_2022}, which are independent of leakage terms between different polarisation feeds. For these reasons, closure quantities have been integral to calibration and synthesis imaging in radio astronomy \citep[e.g][]{Rogers_1974, Readhead_1978, Cornwell_1999}. 

\section{Closure Invariant Formalism}\label{sec:CI-formalism}
\citet{Thyagarajan_2022_CI} and \citet{Samuel_2022} recently presented a generalised formalism of ``closure invariants'' for co-polar and polarimetric interferometric measurements, respectively. In this work, we focus on co-polar closure invariants developed in \citet{Thyagarajan_2022_CI}. 
Within the Abelian gauge theory formalism, closure invariants can be obtained using triangular loops.  An advariant is defined on triangular loops consisting of any pair of array elements $(a,b)$ pinned on a fixed reference vertex indexed at 0,
\begin{equation}
    \mathcal{A}'_{\rm{0ab}} = \mathcal{V}'_{0a}(\mathcal{V}'^{*}_{ab})^{-1}\mathcal{V}'_{b0} = \abs{g_0}^2\mathcal{A}_{\rm{0ab}}\,.
\end{equation}
Here, $\abs{g_0}^2$ is an unknown scaling factor identical on all complex advariants associated with the reference vertex, which can be canceled out by normalising to any one non-zero advariant or any $p$-norm of the advariants to obtain a complete and independent set of closure invariants. Under this formalism, the total number of $N_s^2 - 3N_s + 1$ real-valued closure invariants is identical to the total number of closure phases and amplitudes, and contain equivalent information. 

Although the exact equivalent information is present in both formalisms, closure invariants derived from the Abelian gauge theoretic framework offer several advantages compared to the traditional closure phases and amplitudes \citep{Thyagarajan_2022_CI}. Traditional closure quantities are historically determined on closed triangular and quadrilateral loops, necessitating separate treatment for each, but the complete set of closure invariants can be obtained entirely from independent triads, the simplest non-trivial loop. This provides a unified treatment of closure-based information in co-polar interferometry. Moreover, with the presence of thermal noise and complex gain error, the traditional representation of closure information as amplitudes and phases introduces statistically-dependent noise covariances \citep[e.g.][]{Blackburn_2020}, which are often disregarded. Meanwhile, even though closure invariants are neither Gaussian nor uncorrelated in general, covariances induced by choice of coordinate system can be avoided when representing closure invariants by their real and imaginary components. Additional detail is presented in \citet{Thyagarajan_2022_CI} and the polarimetric extension in \citet{Samuel_2022}.

\section{Imaging Methods} \label{sec:imaging}

In this section, we briefly describe how VLBI data products are utilised in the multitude of approaches to image reconstruction: \clean\ \citep{Hogbom_1974_CLEAN} and its modern improved variants \citep[e.g.][]{Cornwell_2008_CLEAN, Offringa_2014_WSCLEAN}, regularised maximum-likelihood methods methods \citep[e.g.][]{Chael_2018_ehtim, Mueller_2022_doghit, Mus_2024_PSO}, Bayesian methods \citep[e.g.][]{Arras_2019, Tiede_2022_Comrade}, and machine learning methods \citep[e.g.][]{Schmidt_2022, Sun_2022_adeep}. VLBI Imaging is an ill-posed inverse problem. The forward transformation from a intensity distribution on the sky to a set of observables is deterministic for a fixed interferometric array, in principle, but the reverse transformation has innumerable viable solutions due to the sparse aperture coverage.

\subsection{Current approaches}

\clean\ \citep{Hogbom_1974_CLEAN} confronts the imaging problem with an iterative greedy matching strategy. In radio astronomy, it is the most successful and influential deconvolution algorithm, and its modern variants \citep[e.g.][]{Wakker_1988, Cornwell_2008_CLEAN, Offringa_2014_WSCLEAN} remain extensively used today. Briefly, \clean\ models sources in the field as point sources, iteratively subtracting the scaled beam response until the algorithm has met a termination criteria, where it is determined that no significant sources remain present in the image. The final image reconstruction is a model composed of many point sources convolved with the clean restoring beam. The general assumption that the observed emission is consistent with point sources can become inadequate for modelling single objects at high resolution, such as event-horizon scale structure. Modern variants can model the sky as kernels at different scales \citep[e.g. multiscale-\clean;][]{Cornwell_2008_CLEAN} and are better suited to extended structure.  

In contrast to direct deconvolution, forward-modelling methods update image parameters via optimisation algorithms on an objective function. A class of regularised maximum-likelihood methods \citep[RML; e.g.][]{Ikeda_2016, Akiyama_2017_polarimetric, Akiyama_2017_imaging, Chael_2018_ehtim, Blackburn_2020}, which build off from maximum entropy methods \citep[e.g.][]{Frieden_1972, Narayan_1986}, minimise the weighted sum of data-fidelity and regularisation terms to obtain viable solutions with desired characteristics imposed by the choice and weighting of the regularisation terms. Regularisers can include entropy \citep[e.g.][]{Frieden_1972, Gull_1978, Narayan_1986}, the sparsity-promoting $l1$-norm \citep{Honma_2014}, isotropic total variation \citep{RUDIN1992259}, total squared variation \citep{Kuramochi_2018}, total image flux density, and image centroid position. By minimising an objective function containing only closure data terms, some forward-modelling algorithms, such as \ehtim\footnote{\href{https://github.com/achael/eht-imaging}{https://github.com/achael/eht-imaging}}, have shown that the closure quantities can be leveraged directly for imaging. However, the global objective function landscape, including a comprehensive hyperparameter search, can be highly complex and it is impossible to know the optimal weighting scheme for any image reconstruction problem a priori. Recently, \citet{Muller_2023_MOEAD} and \citet{Mus_2024_PSO} proposed a solution by utilising multiobjective optimisation techniques to perform global searches across the hyperparameter surface and find sets of locally optimal models. 

While other methods can struggle with presenting reliable uncertainties, Bayesian imaging methods \citep[e.g.][]{Arras_2019, Broderick_2020_Themis, Tiede_2022_Comrade, Roth_2024, Liaudat_2024} formulate the imaging problem as one that is tractable by Bayesian inference methods in order to obtain a posterior distribution of the image conditioned on the available data. These methods require a sophisticated description of the data likelihood, which encapsulates all of the information about the measurement device and observing process, including corruption terms. As such, existing methods have, for practical reasons of reducing the compute time and requirements, opted to involve various simplifications in the noise, sky, or prior models. Nevertheless, there is no fundamental reason that more accurate models could not be incorporated into the Bayesian framework. 

\subsection{Machine learning approach}
Recently, machine and deep learning techniques have earned widespread adoption within the field of astronomy across various domains \citep{Longo_2019, Huertas_2023}, driven by the proliferation of large-scale datasets. Notably, deep learning methods have demonstrated remarkable performance on inverse problems, such as image or video denoising and super-resolution \citep[e.g.][]{Zhang_2017, Rombach_2021_StableDiff, Donike_2025}. In the radio interferometric imaging inverse problem, a diverse array of machine learning methods have already been employed for both image reconstruction or classification, including convolutional neural networks \citep[e.g.][]{Sureau_2020, Nammour_2022, Schmidt_2022, Chiche_2023, Terris_2023}, normalising flow \citep[e.g.][]{Sun_2022_adeep, Feng_2024}, denoising diffusion \citep[e.g.][]{Drozdova_2024, Feng_2024}, classifiers \citep[e.g.][]{Rustige_2023, Thyagarajan_2024_Lucas}, adversarial networks \citep[e.g.][]{Geyer_2023, Rustige_2023}, and transformers \citep[e.g.][]{Lai_2025_DIReCT}. 

In this section, we explore the Deep learning Image Reconstruction with Closure Terms \citep[\direct;][]{Lai_2025_DIReCT} method, which employs the transformer and its attention mechanism \citep{Vaswani_2017} to leverage closure invariants derived from the \citet{Thyagarajan_2022_CI} formalism for direct image reconstruction. Then, we discuss a diffusion-based deep generative model that builds on the work of \direct\ to model the distribution of images consistent with both the training dataset and the provided closure invariants. We also include a convolutional neural network which learns the optimal compression for the set of sampled images and ensures concordance between the measured and reconstructed closure invariants.

\subsubsection{\direct}

The Deep learning Image Reconstruction with Closure Terms \citep[\direct;][]{Lai_2025_DIReCT} model is a deep neural network approach designed to reconstruct source morphology from its closure invariants observed through a VLBI array, such as the Event Horizon Telescope \citep{Doeleman_2009}. The machine learning architecture is composed of a convolutional autoencoder \citep[e.g.][]{KRAMER1992313} with skip connections \citep[e.g.][]{He_2016_RESNET}, and a transformer encoder attachment \citep[e.g.][]{Vaswani_2017, Dosovitskiy_2020_ViT}. The purpose of the transformer is to predict the latent features from the image encoder given only measurable closure invariants as input. 

Training \direct\ involved four loss functions: a morphological cross-entropy classification loss, mean squared error between two sets of latent variables, and two independent image fidelity loss terms associated with the decoded output from the image encoder and the transformer. Each of the loss terms were weighted according to a predefined schedule. After convergence, the resulting model was tested with a variety of source morphologies; the results of which were comparable to deconvolution and forward-modelling imaging algorithms in image fidelity metrics. \direct\ was also tested against corruptions by applying synthetic thermal noise to the input observables, and it achieved a reconstruction fidelity score of $\gtrsim90\%$ down to a signal-to-noise ratio of 10 on the closure invariants. Despite its success, the results from \direct\ were deterministic, rendering it difficult to interpret the confidence of any individual image reconstruction, where the ground truth is unknown. Moreover, \direct\ was not configured to directly minimise the loss on the data terms, namely, the closure invariants.

\subsubsection{\gendirect}

In this section, we introduce the novel generative \direct\ model \citep[\gendirect;][]{GenDIReCT_Zenodo} for VLBI imaging, which has two distinct components. The first part of the neural network architecture is based on textually-conditioned image generation \citep[e.g.][]{Rombach_2021_StableDiff} with denoising diffusion probabilistic models \citep{DDPM_ho_2020}, which has emerged as the new state-of-the-art in image generation. As diffusion models effectively capture non-linear relationships and excel at modelling complex data distributions, we use a latent diffusion model to learn the conditional distribution of encoded morphologies from a training dataset consistent with an input set of closure invariants. The second component of the model is a standard multi-layer CNN which is designed to compress a set of sampled images from the diffusion model into a representative image reconstruction. In this section, we describe both parts of the \gendirect\ model in detail.

The process of diffusion is a forward Markov chain process that iteratively adds noise to an image until it is consistent with pure isotropic Gaussian noise. The machine learning architecture learns how to conditionally reverse the noising process at each step until an image can be randomly sampled from noise. In \ref{appendix:diffobj}, we provide a full description of the objective function in conditional denoising diffusion models and additional details of our implementation. In brief, the network is optimised to model $p_{\theta}(x|y) \sim p_{\rm{data}}(x|y)$, the probability distribution of the images $x$ conditioned on closure invariants $y$ with learned model parameters $\theta$. This can be achieved by taking gradient descent steps based on the objective function, defined as 
\begin{equation}
    \mathcal{L}_{\rm{diffusion}} = \mathbbm{E}_{x_t, y, t, \epsilon_t}\left[\left\Vert\epsilon_t - \epsilon_\theta\left(x_t, t, y\right)\right\Vert^2\right], \label{eq:gendirect-loss}
\end{equation}
where $\mathbbm{E}$ is the expectation operator for the difference between the error $\epsilon$ applied to the image $x$ at timestep $t$ and the error predicted by the network $\epsilon_\theta$ parameterised by $\theta$. The probability distribution, $p_{\rm{data}}$, modelled by the network is supplied by the training dataset. 

\begin{figure*}
	\includegraphics[width=0.95\textwidth]{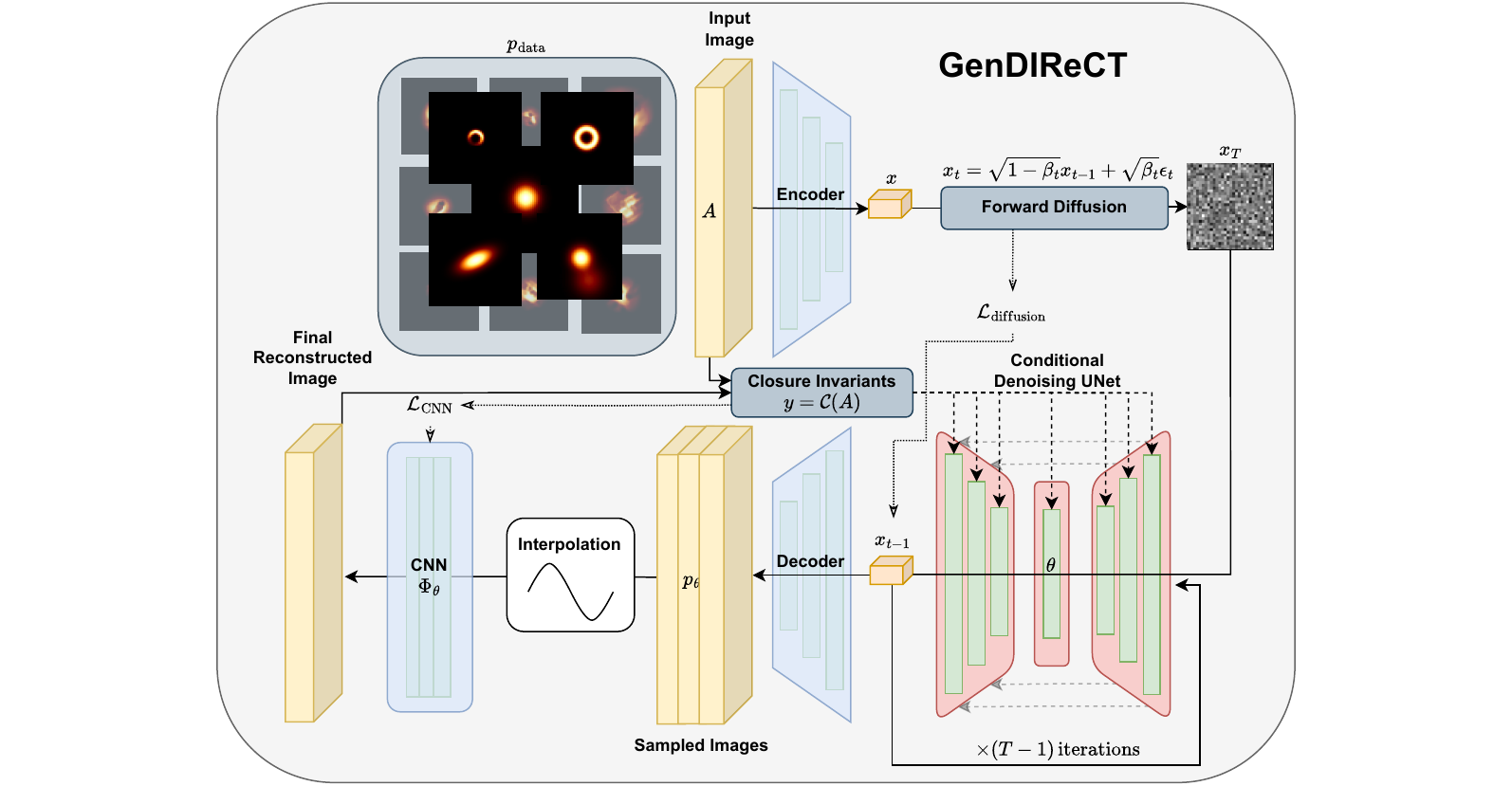}
    \caption[]{Diagram of the \gendirect\ architecture. In the diffusion network, closure invariants measured for images in the training dataset are used to condition the denoising UNet and furthermore, as targets for the convolutional network. The conditional UNet is trained to reverse the diffusion process by taking gradient descent steps on the $\mathcal{L}_{\gendirect}$ objective function, defined in Equation~(\ref{eq:gendirect-loss}) The result is a sample of images, $p_\theta(x|y)\sim p_{\rm data}(x|y)$, which are inputs for the convolutional neural network. The CNN learns the optimal compression for the set of the sampled images on the image axis based on the loss function defined in Equation~(\ref{eq:cnn-loss}), which ensures that the final reconstructed image is consistent with the input closure invariants.}
    \label{fig:gendirect-architecture}
\end{figure*}

In practice, rather than applying diffusion on the image directly, we leverage the convolutional autoencoder architecture of the original \direct\ model \citep[see Figure 1 of][]{Lai_2025_DIReCT} to project images into lower-dimensional latent features. Therefore, diffusion is applied to encoded latent features, which can subsequently be decoded into images. This latent diffusion approach is introduced and described in \citet{Rombach_2021_StableDiff}. With this approach, the diffusion model optimises the distributional objective in the latent domain, allowing one to generate plausible sets of latents conditioned on a given set of observed closure invariants. We note that alternatively, one could replace the autoencoder in the \gendirect\ architecture with one that is unrelated to \direct; however, in our evaluation, we have observed that the performance of the diffusion model is significantly deteriorated, implying that in the process of training \direct, the latent features are arranged in a manner that is conducive to being predicted by closure invariants. It is not necessary for the \direct\ autoencoder to be trained with exactly the same VLBI array and observation synthesis as the one used for \gendirect. We employ a UNet \citep{Ronneberger_2015_Unet} architecture for the diffusion model and a visual representation is presented in Figure \ref{fig:gendirect-architecture}, where green bars represent convolutional layers. The set of output images sampled from the diffusion model, $p_{\theta}$, is then fed into the second component of the \gendirect\ system. 

The second part of \gendirect\ is a convolutional neural network (CNN) which simultaneously downsamples input images and compresses the information from all provided images into one final reconstruction. The advantage of the CNN is that it is less restricted by the types of morphologies present in the selected training dataset, enabling its reconstruction to outperform all of the individual diffusion-sampled images and other averaging or alternative aggregation strategies, especially on unseen datasets. Additionally, we can explicitly optimise for the selected data metrics. In this work, the criteria for the optimal compression is based on the $\chi^2$ likelihood function, as 
\begin{equation}
    \mathcal{L}_{\rm{CNN}} = \sum_{i}^{N_{\rm ci}}\left[\frac{\left(\mathcal{C}(\Phi_\theta)_i - \mathcal{C}({A})_i\right)^2}{\sigma_i^2}\right]\,, \label{eq:cnn-loss}
\end{equation}
where $\mathcal{C}$ is the function that maps an image $A$ to its closure invariants, indexed by $i$, given a particular observation arrangement, $\Phi_\theta$ represents the CNN with parameters $\theta$, and $N_{\rm ci}$ is the total number of closure invariants. The normalisation quantity $\sigma_i$ is the standard deviation uncertainty of the $i$-th closure invariant, which is sensitive to the noise model and observation details. We note that while we choose to minimise the $\chi^2$, other loss functions (i.e. L1 or mean squared error loss) and other aggregation operations (i.e. mean rather than sum) can be equally valid. After sampling images from the $p_\theta$ distribution learned by the diffusion model, we perform a smooth cubic interpolation to upscale each image in the sample before passing them into the CNN to learn the optimal compression based on the Equation~(\ref{eq:cnn-loss}) objective function. We find that introducing such redundancies from the interpolation help manifest smoother final image reconstructions through the CNN. The data-aware CNN is comprised of three layers of 2-D convolutional operations, interspersed with non-linear activation functions with a final activation function placed before the output to enforce image positivity. We find that beyond a modest depth, alterations to the design of the architecture do not significantly influence the resulting reconstruction. It's also possible to use the interpolation and CNN to produce image reconstructions at a slightly different field of view or resolution than the one used to train the diffusion model. However, because the CNN output is sensitive to its input images sampled from the diffusion model with a fixed field-of-view and resolution, the dimensions of the final image reconstruction should be kept relatively consistent with the input images, in order to remain aligned with the spatial and frequency scales of the diffusion output. 

\subsubsection{Training \gendirect} \label{sec:training}

\begin{figure*}
	\includegraphics[width=0.9\textwidth]{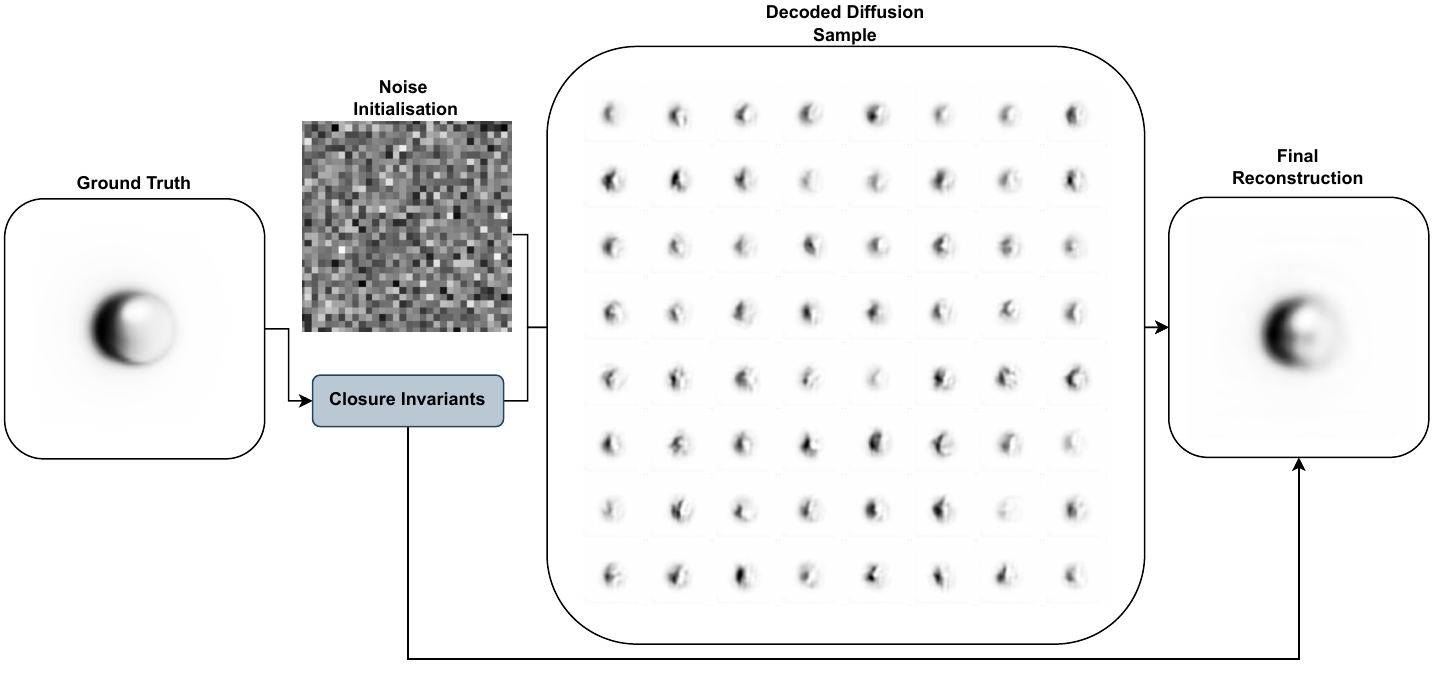}
    \caption[]{Illustration of the output from each layer of the \gendirect\ architecture and typical imaging procedure. The illustration utilises a simulated image of Sgr A*, which is not part of the training dataset. Random noise is used to initialise the denoising UNet conditioned on closure invariants observed from the ground truth image to create a sample of latent information, which can be decoded into images. We show 64 images sampled from the diffusion model, all of which show a crescent-like structure. The final image is reconstructed by the CNN by learning the optimal compression of the diffusion sample. }
    \label{fig:gendirect-outputs}
\end{figure*}

The two components of the \gendirect\ architecture are trained independently with different philosophies. The diffusion model is trained in a supervised manner, by utilising a training dataset containing a wide variety of possible morphologies. The CNN, on the other hand, is unsupervised and its input, a sample of images from the diffusion model, remains unchanged throughout its entire training process. Moreover, the trained CNN is not generally useful as it is designed to be highly specific to the input data and closure invariants. As such, it will need to be retrained for each independent reconstruction.  

The training dataset of \gendirect's diffusion model is identical to that of \direct\ and consists of a variety of extended sources, which include Gaussians, disks, rings, ellipses, and $m$-rings of first and second order \citep{Roelofs_2023_mring}. However, the majority of the dataset is comprised of the CIFAR-10 dataset \citep{Krizhevsky09_CIFAR10}, a set of 60,000 images representing ten natural or man-made object classes (e.g. dogs, horses, planes, ships, etc). As the original resolution of CIFAR-10 is merely $32\times32$ pixels, we use a super-resolution $128\times128$ pixels version of CIFAR-10 resized by a \textsc{Pascal}-based deep learning neural network \citep{cai_neural_api_2021_5810077}. Random augmentations were applied to all images during training to further expand the variety of images encountered by the model. The augmentations included image flipping and affine transformations such as rotating, rescaling, and shearing. The CIFAR-10 images are augmented with an additional radial tapering mask, defined as an element-wise multiplication operation between the original image with a disk blurred by a Gaussian kernel. The parameters of the disk and its blurring kernel are randomised during training. 

We train the diffusion model for 200 epochs with the \textsc{Adam} optimiser \citep{Kingma_2014} and a fixed learning rate of 0.001. Our variance schedule is cosine-based \citep{Nichol_2021_ImprovedDDPM}, which is designed to preserve information for longer time compared to a linear-based schedule \citep[e.g.][]{DDPM_ho_2020}. We also select $T=1000$ diffusion timesteps. The diffusion architecture is kept relatively compact, with only $\sim 4$M trainable parameters; however, this number depends sensitively on the observation setup, the volume of data, and the specific design of the denoising UNet. A different setting (e.g. aperture coverage or field of view) can require a more complex architecture to achieve ideal performance, but a full exploration of the deep neural network design exceeds the scope of this work. During training, we measure a running average of the last 100 loss values from the diffusion model and save the model with the minimum loss. The total wall-clock time for training a diffusion model on the described dataset is $\sim 10$ hours on an NVIDIA H100 GPU, sensitive to the aperture coverage and hence, the data volume. 

In contrast to the diffusion model, the CNN is trained in an unsupervised manner by passing the interpolated sample of generated images into the CNN at every epoch. The weights and biases of the CNN are optimised for 2000 epochs with an initial learning rate of 0.001, decaying by a factor of $0.999^n$, where $n$ is the epoch counter, or until convergence. During validation, we discovered that initialising the parameters of the CNN by first optimising the network with the $L_1$ loss for 1000 epochs can assist the initial convergence of the model, thereby influencing the final reconstruction. As such, all results presented in Section \ref{sec:results} and \ref{sec:discussion} adopt the $L_1$ loss for the first 1000 epochs before minimising $\chi^2$. 

Identically to the diffusion model, we use the \textsc{Adam} optimiser \citep{Kingma_2014} for the CNN. The total number of trainable parameters is $\sim10$M and the compute time for training the CNN is on the order of a few minutes on the NVIDIA H100 GPU, subject to the data volume. Like the diffusion model, the dominant bottleneck is the compute time allocated to measuring the closure invariants from synthetic observation of the reconstructed image.

\section{Results} \label{sec:results}

In this work, we adopt the array configuration, observing sequence, and subsequent aperture coverage of one of the datasets in the ngEHT Analysis Challenges \citep[][hereafter \citetalias{Roelofs2023_ngeht-challenge}]{Roelofs2023_ngeht-challenge}. The analysis challenges are collections of publicly available synthetic VLBI data, designed to promote rapid development of imaging techniques and algorithms, especially pertaining to anticipated capability upgrades with the ngEHT array. We discuss the performance of our model on these datasets in Section \ref{sec:ngeht-challenge}. Here, we briefly summarise the aperture coverage in one of the datasets, which we adopt for our model. 

We use the \ehtim\ software package presented in \citet{Chael_2018_ehtim}, hereafter referred to as \citetalias{Chael_2018_ehtim}, to synthesise 230 GHz observations with a receiving bandwidth of 8 GHz for non-variable sources placed at the location of M87, RA = $\rm{12^h30^m49.42^s}$ and Dec = $\rm{+12^{\circ}23'28.04''\, (J2000)}$. The array consists of the 8 stations comprising the original EHT2017 VLBI system \citep{EHT_2019_Array}: the Atacama Large Millimeter/submillimeter Array \citep[ALMA;][]{Wootten_2009_ALMA, Goddi_2019_ALMA}, the Atacama Pathfinder Experiment telescope \citep[APEX;][]{Gusten_2006_APEX}, the Large Millimeter Telescope \citep[LMT;][]{Hughes_2010_LMT}, the Pico Veleta IRAM 30 m telescope \citep[PV;][]{Greve_1995_PV}, the Submillimeter Telescope Observatory \citep[SMT;][]{Baars_1999_SMT}, the James Clerk Maxwell Telescope (JCMT), the Submillimeter Array \citep[SMA;][]{Ho_2004_SMA}, and the South Pole Telescope \citep[SPT;][]{Carlstrom_2011_SPT, Kim_2018_SPT}. Also included are three additional stations which were selected based on their favourable geographical location for optimal aperture coverage \citep{Raymond_2021}: the Greenland Telescope \citep[GLT;][]{Chen_2023_GLT}, Kitt Peak 12m telescope \citep[KP;][]{Ziurys_2016_KP}, and the Plateau de Bure IRAM Northern Extended Millimeter Array \citep[PDB;][]{Akiyama_2017_imaging}. Henceforth, the full array, consisting of 11 stations, is referred to as the EHT2022 system \citepalias{Roelofs2023_ngeht-challenge}.

Following the challenge dataset, observations are synthesised from the array with 10 minute integrations interleaved with 10 minute breaks over a period of 24 hours beginning on $\rm{MJD} = 0$, 1858-11-17 midnight UTC. Within each scan, data is captured at a 10s cadence. The dimensions of the nearly circular effective clean beam is $\sim 24\,\mu$as$\times 23\,\mu$as and the aperture coverage is illustrated on the top-left panel of Figure~3 in \citetalias{Roelofs2023_ngeht-challenge}. In the absence of any aggregation operations, the total number of real-valued closure invariants is 76700. However, in order to minimise thermal noise and keep the dataset at a manageable volume, we choose to average the visibility data within each 10-minute scan. The compressed dataset consists of 1294 closure invariants for each observation, 1248 of which are independent. However, we retain all closure invariants for the neural network to learn from the positional encoding for every antenna triad. In \ref{appendix:avg-ci}, we investigate the consequences of averaging closure invariants relative to averaging visibilities in the presence of thermal noise and multiplicative corruptions, concluding that while the former is often quantitatively superior in the presence of time-dependent noise, it's unlikely that an enhancement in reconstruction fidelity would be detectable in this study.

Using the fixed aperture coverage, we train the diffusion model for 200 epochs as described in Section \ref{sec:training} and the model with the lowest running mean loss is saved. To generate an image reconstruction, the conditional denoising UNet is initialised with random Gaussian-distributed noise and closure invariants from an unknown image are concatenated with features passed between convolution layers within the UNet. The resulting latent representation after $T=1000$ timesteps is decoded into an image. We generate $N=1024$ images to serve as the input for the CNN model to create the final image reconstruction while minimising the closure invariants loss term. Figure \ref{fig:gendirect-outputs} illustrates the output of each layer in the \gendirect\ architecture for an example simulated image of Sgr A*, which is not present in the training dataset. We show a grid of 64 images sampled from the diffusion model conditioned on closure invariants and the final reconstruction from the CNN. All of the individual images sampled from diffusion consistently depict a crescent-like morphology, indicative of a reconstruction with relatively high confidence. 

\subsection{Evaluation metrics} \label{sec:metrics}
We quantify the reconstructed image fidelity with the maximum normalised cross-correlation metric, $\rho_{\rm{NX}}$, as \citepalias{Chael_2018_ehtim, Roelofs2023_ngeht-challenge},
\begin{equation}
    \rho_{\rm{NX}}(A_{\rm rec}, T) = \frac{1}{M}\max\abs{{\mathcal{F}^{-1}\{\mathcal{F}\{\hat{A_{\rm rec}}\}\mathcal{F}\{\hat{T}\}^*\}}}\,,
\end{equation}
where forward and inverse Fourier transforms are represented by $\mathcal{F}$ and $\mathcal{F}^{-1}$, respectively, and the operation $\hat{I} = (I - \bar{I})/\sigma_I$ is any image $I$ normalised by its mean and standard deviation. The total number of pixels is represented by $M$, normalising the $\rho_{\rm{NX}}$ where a unit score denotes a perfect reconstruction. In contrast to pixel-by-pixel similarity metrics, the $\rho_{\rm{NX}}$ is insensitive to the absolute position and overall flux of the source, which prevents different reconstructions from being unduly penalised by information that is not constrained by closure quantities. Unless specified otherwise, we measure the $\rho_{\rm{NX}}$ of any reconstructed image, denoted by $A_{\rm rec}$, with respect to an unaltered ground truth image, denoted by $T$. Additionally, we evaluate the goodness-of-fit to data with the reduced $\chi^2$ metric, which is commonly used in other studies to evaluate data adherence in the context of image reconstruction, despite non-linearities in the modelling approach. The reduced $\chi^2$ on closure invariants, denoted by $\chi^2_{\rm{ci}}$, is defined as the sum of error-normalised residuals previously defined in Equation~(\ref{eq:cnn-loss}), divided by the number of independent closure invariants which is the substitute for the degrees of freedom.

Both $\rho_{\rm NX}$ and $\chi^2_{\rm ci}$ operate on the final single-image reconstruction output from the CNN. However, the generative diffusion component of \gendirect\ can produce a variety of intermediate images given a single set of closure invariants. In order to quantify the performance of the diffusion component, we introduce the continuous ranked probability score \citep[CRPS;][]{Hersbach_2000_CRPS}. CRPS is a metric used for measuring the distance between a sample probability distribution function (PDF) and a true PDF. Thus, the CRPS is sensitive to multi-modal posterior probability distributions, a useful property for identifying the presence of morphologically disparate image reconstruction clusters in the output of the diffusion model. In this context, we define the image reconstruction CRPS as,
\begin{equation}
    {\rm CRPS(\textbf{A})} = \frac{1}{M}\sum_{i=1}^{M} \int_{-\infty}^{+\infty} [{\rm CDF}_{i}(A_r) - {\rm CDF}_{i}(A)]^2 dA\,,
\end{equation}
where ${\rm CDF}_{i}(A)$ is the cumulative distribution function of pixel intensity for all diffusion model reconstructed images, $A_r$, and ${\rm CDF}_{i}(A)$ is Heaviside step-function centered on the true intensity of image $A$ at the $i$-th pixel. As a pixel-by-pixel comparison, we shift the images to maximise $\rho_\textrm{NX}$ with the ground truth image prior to estimating the CRPS. Though the numerical value of the CRPS is not interpretable between reconstructions of different sources, we use the CRPS value to measure performance of the diffusion output on the same source under varying levels of additive noise in the following section. Therefore, the CRPS measures the performance of the diffusion model only while $\rho_{\rm{NX}}$ and $\chi^2_{\rm{CI}}$ measures the performance of the final single-image reconstruction after passing the diffusion output into the CNN. We emphasise that neither the CRPS nor $\rho_\textrm{NX}$ are used during model training. They are metrics introduced for evaluating the performance of the trained model.

\begin{figure}
	\includegraphics[width=0.9\textwidth]{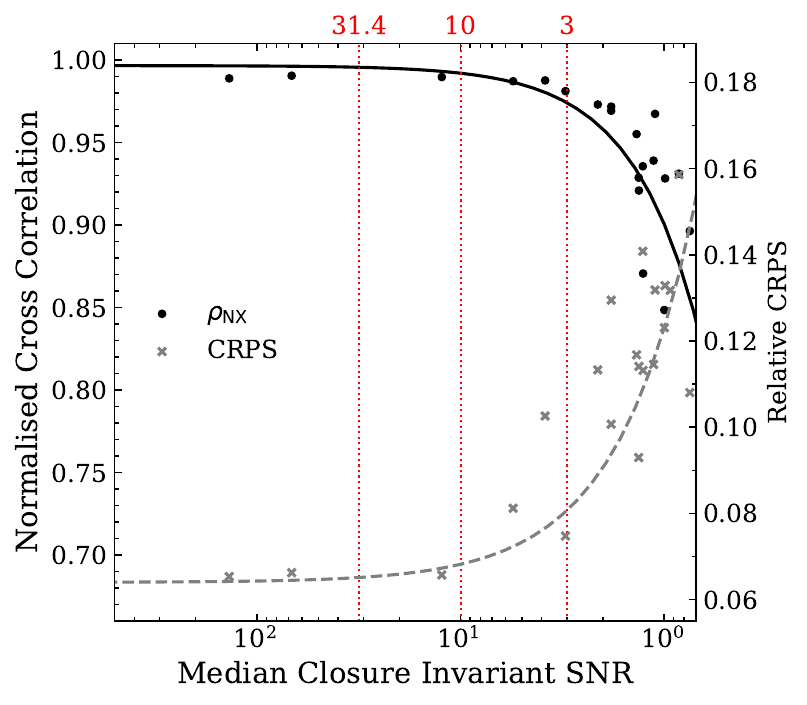}
    \caption[]{Plot of the maximum normalised cross-correlation image fidelity metric, $\rho_{\rm NX}$ (dots), of the final reconstructed image and the relative CRPS (crosses) of the diffusion output as a function of the closure invariants' signal-to-noise ratio on a simulated observation of Sgr A*. Vertical dotted lines mark SNR thresholds of 31.4, 10, and 3, which correspond to the median phase calibrated Stokes I component SNR of the primary M87 EHT dataset \citep{EHT_2019_Data}, a SNR threshold for self-calibration, and a threshold commonly used for low-SNR flagging, respectively.}
    \label{fig:noise_performance}
\end{figure}

\begin{figure*}
	\includegraphics[width=1.0\textwidth]{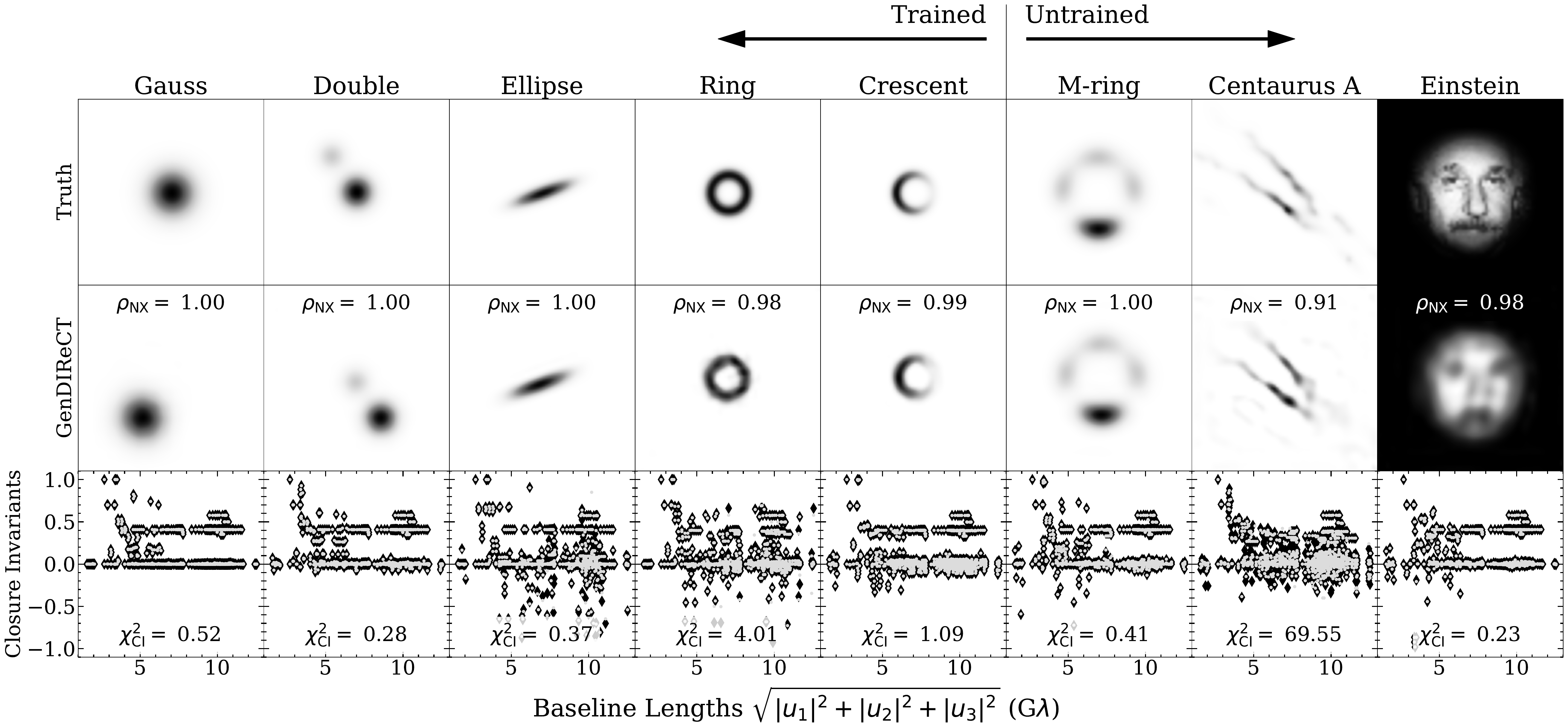}
    \caption[]{Results of the \gendirect\ image reconstruction pipeline on a variety of test images, where the first five basic shapes (Gauss, Double, Ellipse, Ring, and Crescent) are represented in the training dataset, but the latter three ($m$-ring, Centaurus A, and Einstein) are examples of untrained morphologies. The first row presents the ground truth image from which visibilities and subsequently closure invariants are derived. The \gendirect\ middle row presents the final reconstruction from this work's imaging pipeline, alongside the maximum normalised cross-correlation $\rho_{\rm NX}$ image fidelity metric. The bottom row displays the ground truth closure invariants as black diamonds and reconstruction closure invariants as grey points. The final $\chi^2_{\rm CI}$ goodness-of-fit metric is shown. The Einstein model is illustrated with an inverted colormap for enhanced visual clarity.}
    \label{fig:gendirect_results}
\end{figure*}

\subsection{Thermal noise corruption} \label{sec:thermal-noise}
As multiplicative station gain errors and phase corruptions are removed by the closure invariant construction to first-order, we consider how the additive thermal noise affects the signal-to-noise ratio (SNR) of closure invariants and the subsequent reconstruction in this section. With the addition of the thermal noise term, the uncertainty of the closure invariants are no longer agnostic of the overall flux of the source. Thus, we test the robustness of the model for sources normalised to a total flux of 1 Jy and we report the performance with respect to the median closure invariant SNR. We explore the performance of \gendirect\ as a function of SNR by enhancing the system equivalent flux densities (SEFD) of all stations by the same multiplicative factor in order to achieve a desired SNR. Then, we sample an observation at that SNR to produce a reconstruction from the noisy closure invariants.

In Figure \ref{fig:noise_performance}, we illustrate the performance of \gendirect\ on simulated observations of the Sgr A* model displayed in Figure \ref{fig:gendirect-outputs}. We measure the maximum normalised cross-correlation image fidelity metric, $\rho_{\rm NX}$, and the relative CRPS of the diffusion output as functions of the median SNR of closure invariants. In this idealised scenario, \gendirect's performance on the Sgr A* model asymptotically approaches $\rho_{\rm NX} \approx 0.99$ at high SNR and its resilience against noise is demonstrated by the consistent performance down to $\rm{SNR} \approx 3$, only dropping to $\rho_{\rm NX} \lesssim 0.90$ near $\rm{SNR} \lesssim 1$. The performance of the diffusion model, as measured by the CRPS, follows a similar general trend, implying a similar response to noise between the diffusion component of \gendirect\ and the output of the CNN. If this performance is maintained in non-ideal scenarios, it implies that under the standard SEFDs of the EHT2022 array, \gendirect\ can be used to produce good quality reconstructions of sources with total fluxes several times fainter than M87, and potentially as low as 0.1 Jy. While this was achieved with \gendirect\ trained on noiseless synthetic observations, it's conceivable that incorporating the noise response of closure invariants during training can assist the model in generating predictions with even greater resilience to thermal noise.

As the closure invariants construction is not necessarily unique \citep{Thyagarajan_2022_CI}, different forms of the closure invariants can influence \gendirect's response to noisy corruptions due to heterogeneity in the array. \citet{Blackburn_2020} demonstrated that when the array's sensitivity is dominated by a single station, constructing closure phases centered around baselines of that station would minimise the closure phase covariances. Closure invariants would similarly benefit from strategic selection of reference antenna, which define the specific manifestation of these invariants. We confirm that varying the reference baselines does not influence noiseless reconstructions, as the information content is identical. However, we caution that the robustness of the reconstruction to noisy corruptions will depend on these decisions. 

\subsection{Trained and untrained morphologies} \label{sec:untrained_morph}
In this section, we present the product of the \gendirect\ reconstruction pipeline in Figure \ref{fig:gendirect_results} for both trained and untrained morphologies, normalised to 1 Jy. The trained morphological classes are the 2-D Gaussian, double Gaussian, ellipse, ring, and crescent. The untrained morphologies include a fourth-order $m$-ring \citep{Roelofs_2023_mring}, Centaurus A model \citep{Janssen_2021_CenA}, and Einstein's face. In order to adapt Einstein's face for our model, we have augmented the image with a radial tapering mask, similar to those applied to the CIFAR-10 dataset. The field-of-view of all images is fixed to $225\,\mu$as$\,\times\,225\,\mu$as and the image dimensions are $64\times64$ pixels. 

We achieve near-flawless performance on image fidelity metrics for all trained morphologies, with $\rho_{\rm{NX}} \gtrsim 0.98$. Note that any departure in $\rho_{\rm{NX}}$ from unity under $0.005$ has been rounded up and a shift in the position of some reconstructions, most evident for the Gaussian and double Gaussian models, are consequences of ambiguity in the absolute position of the source when closure invariants are the only available data terms. For four of the five trained morphologies, the $\chi^2_{\rm{CI}}$ data metric shows decent performance. The conventional interpretation of reduced $\chi^2$ centered around the ground truth value of unity would not be applicable in this context, as closure invariants exhibit non-trivial covariances between each other and the construction of the closure invariants is a significantly non-linear process. Consequently, the effective degrees of freedom relevant to the reduced $\chi^2$ is not straightforward. Nevertheless, it is possible to control the final reconstruction $\chi^2_{\rm{CI}}$ value and minimise instances of overfitting to noise by implementing convergence criteria during CNN training. The ring model appears as an exception where it is faithfully reconstructed in the image domain, but exhibits noticeably weaker performance on $\chi^2_{\rm{CI}}$. Indeed, while the $\rho_{\rm{NX}}$ score is good, non-uniformities in the ring surface brightness can be observed along its circumference. By inspecting the output of the reverse diffusion process, we observe that there is minimal morphological diversity in the sampled images from the denoising UNet model. While indicative of a reconstruction with high confidence, the final image produced by the CNN retains characteristics of the diffusion sample, notably its difficulty in adhering to closure invariants. By applying small augmentations to the generated sample to regenerate the generative sample diversity, we can produce an alternative reconstruction with a more uniform surface brightness and improved performance on both metrics ($\rho_{\rm{NX}} = 0.99$ and $\chi^2_{\rm{CI}} = 0.45$, not shown here). Hence, we find that a suitable diversity of images for the CNN to compress is a prerequisite for good performance on data metrics. Therefore, we recommend inspecting the diffusion output to ensure morphological diversity in the sample prior to applying the CNN operation. 

Despite not being represented in the training dataset, the performance of \gendirect\ on the fourth-order $m$-ring is comparable to that of the trained morphologies. However, other untrained morphologies, such as the Centaurus A edge-brightened jet and Einstein's face, are more considerable challenges. Although human faces are not in the CIFAR-10 dataset, Einstein's radially tapered face is accurately reconstructed by \gendirect\ and the adherence to data is also excellent. Facial characteristics such as the eyebrows and eyes, outline of the nose, moustache, facial shape, and brightness asymmetry between the cheeks can all be identified in the reconstructed image. 

The lowest image fidelity reconstruction in this untrained morphology test set is that of Centaurus A, which still achieved $\rho_{\rm{NX}} > 0.9$. There is also a notable degradation in the $\chi^2_{\rm{CI}}$ metric compared to other morphologies. Spurious low luminosity artefacts, which are parallel to the edge-brightened jet, can be observed. These features resemble the artefacts shown in the closure-only reconstructions in Figure 7 of \citetalias{Chael_2018_ehtim}, which are explained as local minima in the regularised maximum likelihood objective function. In our case, the poor performance in $\chi^2_{\rm{CI}}$ is the result of the constrained field-of-view of images in the training dataset, where extended low-luminosity emission is never present near the edge of the image. By applying the identical radial tapering mask as on Einstein's face to Centaurus A and thereby removing these extended features, we are able to recover $\rho_{\rm{NX}} = 0.98$ and $\chi^2_{\rm{CI}} = 0.30$ with no artefacts, which aligns with the performance on other morphologies. The alternative augmented reconstruction of Centaurus A is not shown in the figure. From this, we conclude that as long as much of the overall flux is confined within the fixed field-of-view of the model appropriate for the interferometric array, \gendirect\ would be capable of producing excellent image reconstructions under the constraints of sparse aperture coverage typical of VLBI. 


\begin{figure*}
	\includegraphics[width=0.9\textwidth]{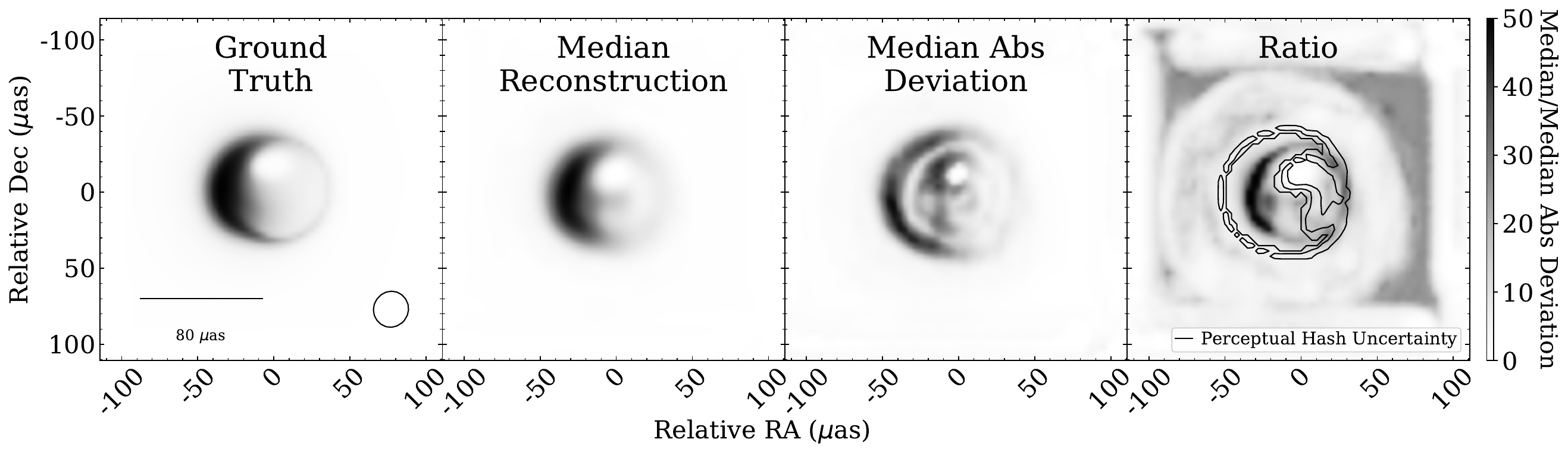}
    \caption[]{From left to right: Ground truth image, median image reconstruction, median absolute deviation image of all reconstructions, and ratio image of the median to the median absolute deviation, which illustrates a `signal-to-noise' ratio of image reconstructions. Large values in the ratio image indicate pixels with low variance relative to the mean pixel intensity. Contours on the ratio image highlight pixels with high perceptual hash variance, which corresponds to higher morphological uncertainty. They are observed to occur on regions of low signal-to-noise' ratio, and thus do not signify an appreciable morphological difference.}
    \label{fig:confidence}
\end{figure*}

\subsection{Reconstruction confidence}
By design, \gendirect\ functions as a generative imaging application, enabling the pipeline to produce different final reconstructions originating from the same input dataset. This property enables us to visualise a distribution in image and data fidelity metrics stemming from the generative variety, as well as produce a standard deviation image by aggregating all reconstructions. To isolate the generative diversity of reconstructions, we synthesise a single noisy observation of the Sgr A* model described in Section \ref{sec:thermal-noise} at standard SEFDs. We artificially decrease the data SNR by scaling the total flux of the source model to 0.5 Jy in order to exaggerate the image reconstruction variation and visualise reconstruction confidence at a lower SNR. We then create a sample of final reconstructions by binning the diffusion-sampled image dataset into 100 bins of 1024 images each, from which one CNN final reconstruction is created for each bin. We illustrate the median and median absolute deviation of all final reconstructions in Figure \ref{fig:confidence}, which is more outlier-resistant than the mean and standard deviation metrics. Because closure invariants are insensitive to the overall flux and position, we normalise the flux of all reconstructions and shift by the maximum cross-correlation prior to computing the median and median absolute deviation. We also measure the `signal-to-noise' ratio between the median reconstruction and its median absolute deviation. Large values in the ratio image indicate pixels with low variance relative to the mean pixel intensity, and therefore higher reconstruction confidence.


From Figure \ref{fig:confidence}, we observe the ground truth, median reconstruction, median absolute deviation, and ratio images, respectively. The median absolute deviation and ratio images both exhibit a thin crescent, which is reconstructed at high confidence. 

Due to instabilities from low variance pixels, the ratio image exhibits spurious features in its periphery, including the appearance of a square-shaped image frame and an enclosed low `signal-to-noise' disk. As both the median flux and absolute deviation are diminutive in these regions, it is difficult to interpret the image morphology reconstruction confidence. Moreover, by normalising all reconstructions to the same overall flux without regard for image structure, we can inadvertently introduce flux differences between perceptually identical image reconstructions. Therefore, we separately define a proxy for image morphology reconstruction confidence by leveraging perceptual hashing methods. 

Perceptual hashing is a family of algorithms designed to quantify image similarity, and it is generally used to identify redundancy in an image dataset, cluster images, or perform reverse image search \citep[e.g.][]{phash_SAMANTA2021203}. While more advanced algorithms create hashes from the discrete cosine or wavelet transformed images, we directly hash our images by thresholding the intensity based on the 90th-percentile of the image's non-zero flux distribution. We then visualise the variance of hashes from the large sample of image reconstructions by plotting the perceptual hash standard deviation as contours on the right-most panel of Figure \ref{fig:confidence}. The contours indicate regions of greater uncertainty. The advantage of the hashing operation over the ratio image is that complex structure is no longer manifested over perceptually insignificant regions, such as near the image periphery. Rather, we can immediately identify a thin boundary across the ring circumference, where variations in the ring size between reconstructions measured at the threshold intensity are expected to remain on the order of one pixel width, which corresponds to $\sim 3.5\,\mu$as. The size of the internal depression also remains consistent to one pixel width. However, the more non-trivial result is at the faint edge of the reconstructed ring, where perceptual differences between reconstructions can be identified where the emission is boosted away from the line-of-sight. These differences in the location of the faint edge between reconstructions manifest as a larger contour-enclosed area, expressing higher perceptual hash uncertainty. However, we observe that the perceptual hash uncertainty contours cover regions of low `signal-to-noise' ratio, and hence they do not represent a significant morphological difference.

\begin{figure*}
	\includegraphics[width=1.0\textwidth]{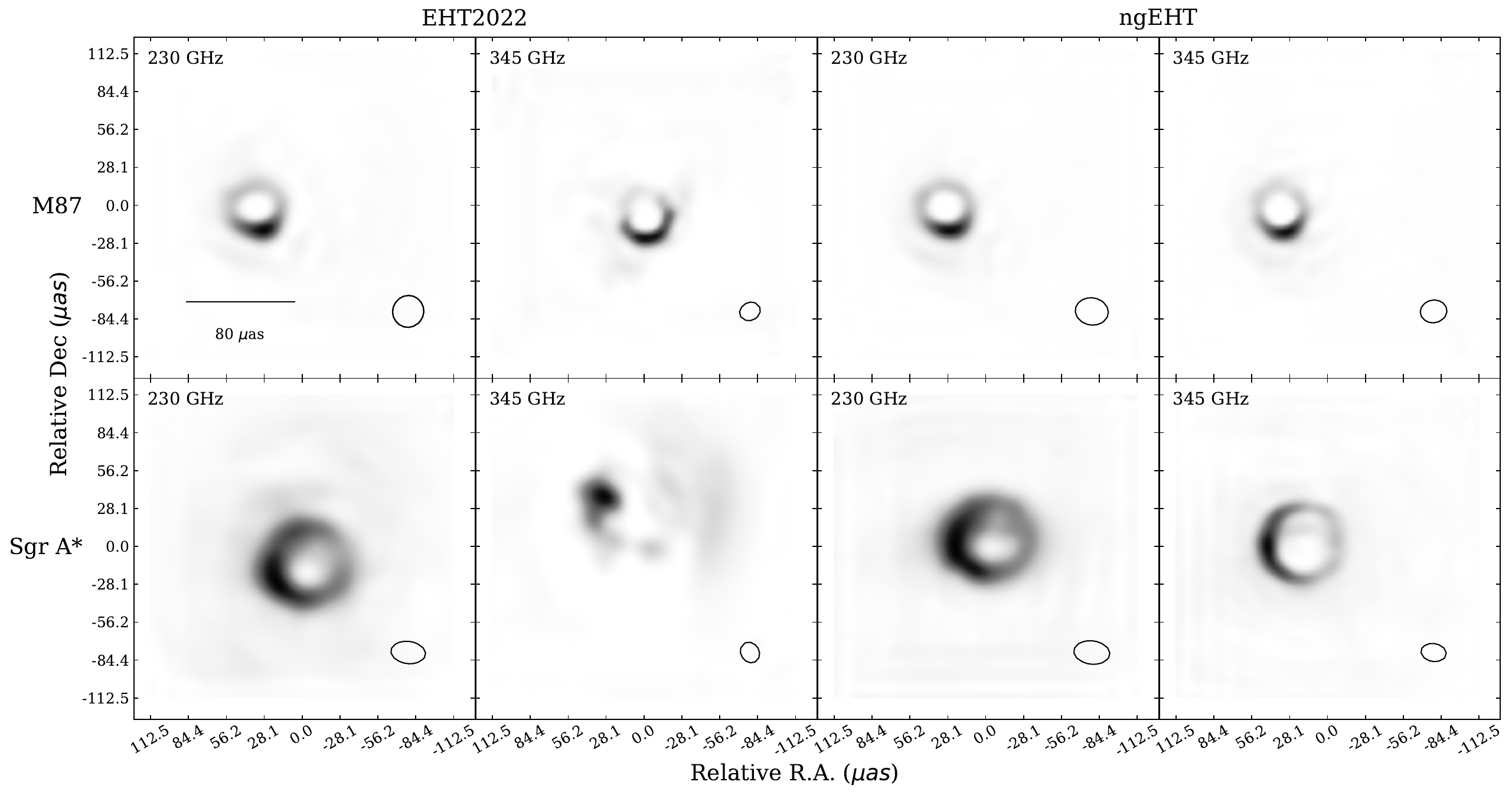}
    \caption[]{\gendirect\ reconstruction for all 8 datasets in the ngEHT Analysis Challenge 1, separated by the observation array (EHT2022 and ngEHT), frequency (230 and 345 GHz), and source (M87 and Sgr A*). In all panels, the effective beam size and shape is illustrated in the bottom right corner. All reconstructions convincingly recover the black hole shadow except the Sgr A* source observed with the EHT2022 array at 345 GHz.}
    \label{fig:ngeht-analysis-challenge}
\end{figure*}

\section{Discussion} \label{sec:discussion}

One of the advantages in the approach we have taken with \gendirect\ is that once we have selected the training dataset, output image properties, data post-processing steps, and a fixed architecture\footnote{Note that while the network architecture is tunable in principle, sufficiently deep networks, when independently optimised for the desired task, are generally insensitive to minute design decisions.}, there remains a limited set of additional hyperparameters that can be tweaked to influence the reconstruction. Unlike alternative imaging techniques, there are no tunable calibration datasets, explicit priors, nor regularisation terms with independent weighting schemes. Rather, the augmented CIFAR-10 dataset and simple geometric models impose an implicit prior, and the diffusion model learns the effective regulariser based on its training. Nevertheless, this procedure can introduce a few biases. For instance, the image interpolation applied prior to the CNN optimisation encourages reconstructions with smoothly varying surface brightness distributions and image positivity is enforced by a final non-linear activation function. The performance of the diffusion model on any particular test source also depends on the presence of structurally similar sources in the training data, although the CNN refinement strategy aids generalisation on unseen morphologies as seen in Figure \ref{fig:gendirect-outputs}. The decision to apply a radial tapering augmentation on the CIFAR-10 dataset encourages the model to favour reconstructing centralised emission sources within its field-of-view. 
The remaining tunable hyperparameters in \gendirect\ are related the cost function of the CNN: whether it is $\chi^2_{CI}$ as we have adopted or other equally valid loss functions. Due to the sparsity of explicit tunable hyperparameters, \gendirect\ provides a relatively simple and easily reproducible imaging tool for VLBI. 

In the previous section, we described a suite of internal tests we developed to evaluate the results of \gendirect\ on noise, unseen morphologies, and reconstruction confidence. We demonstrated the generative \gendirect\ deep learning model can successfully reconstruct general morphologies under the challenging constraints of extreme data sparsity characteristic of the Event Horizon Telescope aperture coverage. Additionally, we find the result is robust against station-based corruptions and thermal noise. However, the model performance is best when much of the overall flux of the source is confined within the fixed field-of-view of the model. While all of the previous tests utilise our own synthetic datasets, in this section, we test \gendirect\ on datasets which use different noise models and critically, are not created by our own aperture synthesis pipeline.

\subsection{ngEHT Analysis Challenge 1} \label{sec:ngeht-challenge}
The ngEHT Analysis Challenges\footnote{\href{https://challenge.ngeht.org/}{https://challenge.ngeht.org/}} are a set of public data challenges motivated by the anticipated performance upgrades the ngEHT would deliver compared to the EHT array. These challenges are designed to encourage rapid development of imaging algorithms and strategies tailored to datasets produced by the ngEHT, particularly by expanding imaging techniques to accommodate both the time and frequency domain. There are four total analysis challenges with increasing complexity, and each of them tests a different component of the image reconstruction inverse problem: 1. total intensity imaging, 2. dynamic imaging, 3. polarimetric imaging, 4. parameter estimation. In this work, we focus on Challenge 1, the total intensity imaging challenge, for which 8 datasets were provided, containing two sources (M87 and Sgr A*) at two frequencies (230 GHz and 345 GHz), synthetically observed with two arrays (EHT2022 and ngEHT). Details of the ground truth simulations, observation sequence to generate the synthetic data, and noise properties are described in \citetalias{Roelofs2023_ngeht-challenge}. Here, we summarise the most pertinent details related to the specific adaptation of our image reconstruction strategy.

The EHT2022 array and observing sequence of all challenge datasets are as described in Section \ref{sec:results}, where 10-minute scans in the challenge dataset are simulated interleaved with 10-minute breaks on a 24-hour track. The ngEHT array is composed of 10 additional stations, but we note that not all stations in both the EHT2022 and ngEHT can observe at 345 GHz, limiting its aperture coverage. Unlike the synthetic observations used to train our model, the data of the challenge datasets include thermal noise estimated from the receiver and atmospheric opacity. Moreover, visibility phases were randomised between scans sourced by station-based multiplicative corruptions. Their effects are removed by the closure invariants construction as long as care is taken that the data are averaged within each scan, where the phases are stable. The total flux of M87 is on the order of 1 Jy and the total flux of Sgr A* is around 3 Jy. 

From its conception, the ngEHT Analysis Challenge 1 has never been a blind challenge, as the input source models were provided to all participants. However, once we fix the cost functions and network architecture as described in Section \ref{sec:imaging}, there are very few hyperparameters in \gendirect\ that can be tweaked to influence the reconstruction. To tackle the ngEHT total intensity analysis challenge, we use the same training dataset filled with non-astronomy augmented CIFAR-10 images described in Section \ref{sec:training}, adopt a $225\,\mu$as$\times225\,\mu$as field-of-view over a $64\times64$ pixel grid, and we choose to average the visibilities within each scan before computing closure invariants. From there, we apply the procedure described in Section \ref{sec:training} to train the \gendirect\ model and perform the image reconstruction using the challenge dataset's data as illustrated in Figure \ref{fig:gendirect-architecture}. 

In addition to the fidelity metrics introduced in Section \ref{sec:metrics}, we consider the traditional closure amplitude and phase $\chi^2$ metrics as well. We compute the $\chi^2$ metrics on closure quantities derived from pre-averaged visibility data in order to perform a direct comparison with other reconstruction strategies. Following \citetalias{Roelofs2023_ngeht-challenge}, we also implement the effective resolution metric, $\theta_{\rm{eff}}$, defined as the FWHM of a 2D circular Gaussian kernel, $\mathcal{G}_{\theta_{\rm eff}}^{\rm 2D}$, convolved with the ground truth image, such that the $\rho_{\rm{NX}}$ of the blurred truth with respect to the unblurred ground truth is identical to the $\rho_{\rm{NX}}$ measured for the reconstructed image. Leveraging the effective resolution, we measure a proxy for the dynamic range, which is defined as \citepalias{Roelofs2023_ngeht-challenge},
\begin{equation}
    \mathcal{D} = \frac{{\rm max}\left(T * \mathcal{G}_{\theta_{\rm eff}}^{\rm 2D}\right)}{\abs{A_{\rm rec} - T * \mathcal{G}_{\theta_{\rm eff}}^{\rm 2D}}}\,,
\end{equation}
where $T$ is the ground truth image and $*$ denotes the convolution operation with a Gaussian kernel characterised by the effective resolution of the reconstructed image, $A_{\rm rec}$. The new array, $\mathcal{D}$, which inherits the dimensions of the image, is the ratio between the brightest pixels of the blurred ground truth image with the residuals between the reconstruction and the blurred ground truth. $\mathcal{D}_{0.1}$, where $D_q = {\rm quantile}\left(\mathcal{D}, q\right)$ is defined by \citetalias{Roelofs2023_ngeht-challenge} as a proxy for the dynamic range to evaluate the performance of distinct reconstruction methods. 

We present the \gendirect\ reconstruction for all 8 datasets in Figure \ref{fig:ngeht-analysis-challenge}. Seven of the eight reconstructed images convincingly recover the black hole shadow and accurately reconstruct its size and shape when compared to the supplied ground truths. For the one exception of the Sgr A* source observed with the EHT2022 array at 345 GHz, the aperture coverage at 345 GHz with the EHT2022 array is too sparse for a high-fidelity reconstruction and no independent submitter presented a successful reconstruction of the black hole shadow with that dataset \citepalias{Roelofs2023_ngeht-challenge}. 

In Table \ref{tab:ngeht-metrics-M87}, we compare the measured evaluation metrics described in Section \ref{sec:metrics} for our M87 reconstructions with those of independent submissions, whose results are presented in Table 1 of \citetalias{Roelofs2023_ngeht-challenge}. As we do not attempt to reconstruct the image in the full $1$ mas field-of-view of the simulation, we compare our reconstruction only to the submissions of Alexander Raymond, Nimesh Patel, and a single reconstruction from TeamIAA \citepalias{Roelofs2023_ngeht-challenge}. These submissions were selected for comparison because limitations had been imposed on the reconstruction field-of-view, similar to our work. Despite using only the closure invariants for image reconstruction, the performance of the \gendirect\ images on most image-related metrics are second only to A. Raymond's reconstructions. We achieve competitive performance on data metrics. Overall, we identify similar trends in our results as those seen by \citetalias{Roelofs2023_ngeht-challenge}. The image reconstruction quality for 345 GHz is generally worse than for 230 GHz due to lower source flux and poorer aperture coverage. However, while the dynamic range of other reconstruction methods typically improve with the enhanced aperture coverage of the ngEHT array, the \gendirect\ dynamic range stays roughly the same. The most probable reason for this is that our reconstructions utilise closure quantities, which are known to have issues with dynamic range \citep[e.g.][]{Chael_2018_ehtim}, whereas the other submissions use the full visibilities. Another confounding factor is that \gendirect\ is trained on an 8-bit image dataset, which imposes limits on the dynamic range of the diffusion reconstruction.

In Table \ref{tab:ngeht-metrics-SGRA}, we compare the evaluation metrics for our Sgr A* reconstructions with those of independent submissions. TeamIAA provided submissions using three different algorithms (\clean, \ehtim, and \smili). As other submissions already use \ehtim, we pick a single reconstruction from either \clean\ or \smili\ with the best $\rho_{\rm NX}$ to present in the table. The dynamic ranges of Sgr A* reconstructions are not reported in Table 1 of \citetalias{Roelofs2023_ngeht-challenge}. Unlike with M87, the \gendirect\ reconstructions rank first in image fidelity metrics for all datasets except the 345 GHz observation with the EHT2022 array, where none of the submissions was able to accurately reconstruct the morphology of the black hole shadow. The improved relative performance can be attributed to the absence of a low surface brightness extended jet, which would extend beyond the fixed field-of-view on which the \gendirect\ model was trained. Rather, much of the overall flux of Sgr A* is confined within \gendirect's field-of-view, which is one of the conditions identified in Section \ref{sec:untrained_morph} for the best performance from the model.

\setlength{\extrarowheight}{3pt}
\begingroup
\begin{table*}
\caption {\label{tab:ngeht-metrics-M87} Reconstruction evaluation metrics for M87 comparing the \gendirect\ reconstruction to submissions of Alexander Raymond and Nimesh Patel, with one reconstruction from TeamIAA \citepalias{Roelofs2023_ngeht-challenge}. The submissions chosen for comparison are based on whether the reconstruction field-of-view had been constrained, similar to this work.} 
\begin{tabular}{lllllcccccc}
\hline \hline
 Source & Array & $\nu$ (GHz) & Submitter & Method & $\chi^2_{\rm{cphase}}$  & $\chi^2_{\rm{lcamp}}$ & $\chi^2_{\rm{ci}}$ & $\rho_{\rm NX}$ & $\theta_{\rm{eff}}$ & $\mathcal{D}_{0.1}$ \\
 \hline
M87 & EHT2022 & 230 & N. Patel & \ehtim & 3.66 & 1159.6 & -- & 0.77 & 21.2 & 418 \\
M87 & EHT2022 & 230 & A. Raymond & \ehtim & 2.28 & 1.77 & -- & 0.90 & 8.0 & 291 \\
\hline
M87 & EHT2022 & 230 & \textbf{This Work} & \gendirect & 1.21 & 1.58 & 1.15 & 0.86 & 10.1 & 494 \\
M87 & EHT2022 & 230 & \textbf{This Work} & \ehtim* & 1.88 & 1.80 & 4.63 & 0.85 & 11.6 & 371 \\
\hline \hline
M87 & EHT2022 & 345 & N. Patel & \ehtim & 1.20 & 7.29 & -- & 0.79 & 16.7 & 734 \\
M87 & EHT2022 & 345 & A. Raymond & \ehtim & 1.19 & 0.62 & -- & 0.88 & 8.2 & 700 \\
M87 & EHT2022 & 345 & TeamIAA & \smili & 1.19 & 0.62 & -- & 0.79 & 16.7 & 645 \\
\hline
M87 & EHT2022 & 345 & \textbf{This Work} & \gendirect & 1.19 & 1.63 & 1.39 & 0.85 & 12.3 & 356 \\
M87 & EHT2022 & 345 & \textbf{This Work} & \ehtim* & 1.20 & 0.61 & 1.76 & 0.88 & 9.0 & 259  \\
\hline \hline
M87 & ngEHT & 230 & N. Patel & \ehtim & 3.50 & 89.74 & -- & 0.83 & 14.6 & 640 \\
M87 & ngEHT & 230 & A. Raymond & \ehtim & 1.65 & 2.14 & -- & 0.92 & 6.2 & 532 \\
\hline
M87 & ngEHT & 230 & \textbf{This Work} & \gendirect & 
1.39 & 1.76 & 1.16 & 0.88 & 8.5 & 493 \\
M87 & ngEHT & 230 & \textbf{This Work} & \ehtim* & 1.95 & 3.51 & 2.64 & 0.86 & 10.1 & 381 \\
\hline \hline
M87 & ngEHT & 345 & N. Patel & \ehtim & 1.20 & 9.99 & -- & 0.79 & 16.7 & 853 \\
M87 & ngEHT & 345 & A. Raymond & \ehtim & 1.17 & 1.00 & -- & 0.91 & 5.7 & 782 \\
\hline
M87 & ngEHT & 345 & \textbf{This Work} & \gendirect & 
1.17 & 1.04 & 1.60 & 0.87 & 9.9 & 397 \\
M87 & ngEHT & 345 & \textbf{This Work} & \ehtim* &  1.18 & 1.02 & 2.09 & 0.87 & 9.9 & 257 \\
\hline \hline
\multicolumn{10}{l}{\footnotesize
*Closure-only version}
\end{tabular}
\end{table*}
\endgroup
\setlength{\extrarowheight}{0pt}

\setlength{\extrarowheight}{3pt}
\begingroup
\begin{table*}
\caption {\label{tab:ngeht-metrics-SGRA} Reconstruction evaluation metrics for Sgr A* comparing the \gendirect\ reconstruction to all other submissions \citepalias{Roelofs2023_ngeht-challenge}. While TeamIAA presented submissions using three different algorithms (\clean, \ehtim, \smili), we choose to display the one submission from either \clean\ or \smili~ with the highest $\rho_{\rm NX}$ for comparison. } 
\begin{tabular}{lllllccccc}
\hline \hline
 Source & Array & $\nu$ (GHz) & Submitter & Method & $\chi^2_{\rm{cphase}}$  & $\chi^2_{\rm{lcamp}}$ & $\chi^2_{\rm{ci}}$ & $\rho_{\rm NX}$ & $\theta_{\rm{eff}}$  \\
 \hline
Sgr A* & EHT2022 & 230 & N. Patel & \ehtim & 6.08 & 347.88 & -- & 0.80 & 45.5 \\
Sgr A* & EHT2022 & 230 & A. Raymond & \ehtim & 3.02 & 8.27 & -- & 0.89 & 25.2 \\
Sgr A* & EHT2022 & 230 & TeamIAA & \clean & 140.97 & 130.20 & -- & 0.90 & 23.4 \\
\hline
Sgr A* & EHT2022 & 230 & \textbf{This Work} & \gendirect & 1.02 & 1.34 & 1.12 & 0.96 & 20.0 \\
Sgr A* & EHT2022 & 230 & \textbf{This Work} & \ehtim* & 1.16 & 1.54 & 26.10 & 0.95 & 25.2  \\
\hline \hline
Sgr A* & EHT2022 & 345 & N. Patel & \ehtim & 1.03 & 20.32 & -- & 0.64 & 61.9 \\
Sgr A* & EHT2022 & 345 & A. Raymond & \ehtim & 1.03 & 0.85 & -- & 0.78 & 26.0 \\
Sgr A* & EHT2022 & 345 & TeamIAA & \clean & 71.44 & 66.33 & -- & 0.79 & 24.5 \\
\hline
Sgr A* & EHT2022 & 345 & \textbf{This Work} & \gendirect & 1.03 & 1.30 & 1.15 & 0.76 & 46.3 \\
Sgr A* & EHT2022 & 345 & \textbf{This Work} & \ehtim* & 1.03 & 0.85 & 1.53 & 0.77 & 43.7 \\
\hline \hline
Sgr A* & ngEHT & 230 & N. Patel & \ehtim & 20.23 & 122.65 & -- & 0.65 & 100.0 \\
Sgr A* & ngEHT & 230 & A. Raymond & \ehtim & 1.14 & 1.87 & -- & 0.93 & 18.1 \\
Sgr A* & ngEHT & 230 & TeamIAA & \smili & 1.40 & 8.81 & -- & 0.95 & 14.3 \\
\hline
Sgr A* & ngEHT & 230 & \textbf{This Work} & \gendirect & 1.14 & 1.79 & 1.40 & 0.98 & 12.5 \\
Sgr A* & ngEHT & 230 & \textbf{This Work} & \ehtim* & 1.17 & 2.22 & 5.14 & 0.97 & 16.3 \\
\hline \hline
Sgr A* & ngEHT & 345 & N. Patel & \ehtim & 2.18 & 15.58 & -- & 0.64 & 61.9 \\
Sgr A* & ngEHT & 345 & A. Raymond & \ehtim & 1.14 & 1.15 & -- & 0.90 & 10.6 \\
Sgr A* & ngEHT & 345 & TeamIAA & \smili & 1.17 & 1.23 & -- & 0.89 & 11.7 \\
\hline
Sgr A* & ngEHT & 345 & \textbf{This Work} & \gendirect & 1.14 & 1.20 & 1.58 & 0.96 & 7.6 \\
Sgr A* & ngEHT & 345 & \textbf{This Work} & \ehtim* & 1.16 & 1.27 & 6.76 & 0.92 & 18.0  \\
\hline \hline
\multicolumn{10}{l}{\footnotesize
*Closure-only version}
\end{tabular}
\end{table*}
\endgroup
\setlength{\extrarowheight}{0pt}

\subsubsection{Comparison with closure-only \ehtim}

By restricting the data terms in the regularised maximum likelihood algorithm, \ehtim, to the (log)-closure amplitudes and closure phases, we can create and discuss closure-only reconstructions for comparison with \gendirect\ on the imaging challenges as part of this work. \citetalias{Chael_2018_ehtim} presented descriptions of their imaging workflows, and the code used in practice for M87 Stokes I imaging on the April 2017 EHT observations is publicly accessible in a GitHub repository\footnote{\href{https://github.com/eventhorizontelescope/2019-D01-02}{https://github.com/eventhorizontelescope/2019-D01-02}}. To achieve the best possible performance from many classical synthesis imaging algorithms, such as \ehtim, it is important to tune and adjust the many hyperparameters of the imaging operation, including the number of iterations, image prior, initialisation, data terms, regularisation terms, and weighting distribution. However, \gendirect\ presents a machine-learning approach for VLBI image reconstruction that achieves its results without any further adjustment for all challenge datasets. Therefore, for this comparison, we do not expend the effort to tune \ehtim\ for each reconstruction as additional tweaking would affect reproducibility and it would be impossible to ensure that the effort is distributed homogeneously. Instead, we describe our generalised adaptation of \ehtim\ for closure-only imaging, inspired by \citetalias{Chael_2018_ehtim} workflows, to perform the imaging task for all 8 challenge datasets under the same general constraints as \gendirect.

First, we fix the reconstruction field-of-view to $225\,\mu$as $\times225\,\mu$as and the image dimensions to $64\times64$, identical to \gendirect. The prior is a $40\,\mu$as centralised 2-D circular Gaussian. We add another Gaussian of identical shape, offset by $40\,\mu$as on both spatial axes and with $1\%$ the total flux of the central Gaussian. The purpose of the offset Gaussian is to break the symmetry and avoid gradient singularities. From there, our adapted imaging procedure consists of three imaging rounds. Each round has a maximum of 100 iterations and adopts the preceding round's output as its prior. Although \citetalias{Chael_2018_ehtim} suggested that down-weighted corrupted visibilities could be included in the initial minimisation steps to aid convergence, we avoid using visibilities altogether and assign closure phases a weighting factor twice that of the log closure amplitudes in the first imaging round as they are more valuable for producing sensible image priors. We use log closure amplitudes over regular closure amplitudes because \citetalias{Chael_2018_ehtim} found that they were a more robust data term in closure-only imaging applications. All closure quantities are generated from 10-minute aggregated visibility data in the challenge datasets. After the first round of imaging, the weighting term of the log closure amplitudes and closure phases are set to be equal in all subsequent rounds.

All imaging rounds use the total squared variation regulariser weighted consistently at 10\% of the closure phase weighting term. We do not use either the flux or centroid regularisers. Between imaging rounds, the intermediate reconstruction is convolved with a circular Gaussian, which is an operation that only affects the closure amplitude data term. The size of the circular Gaussian is set to the measured effective clean beam between the first two imaging rounds and half the dimensions of the effective clean beam between the final two imaging rounds. These blurring operations aid the convergence of the reconstruction and help to remove spurious high-frequency features.

We note that three imaging rounds with interleaved convolution operations may not be sufficient for minimising the supplied data terms, even with down-weighted regularisers. However, aggressively minimising closure terms may not always improve the quality of the image reconstruction on image fidelity metrics, especially when the field-of-view is fixed. Therefore, we have decided to fix the maximum iterations to the three aforementioned imaging rounds and report the outcome, presenting simple and reproducible \ehtim\ results as a reference for \gendirect.

The results of the reference \ehtim\ closure-only reconstructions are presented in Tables \ref{tab:ngeht-metrics-M87} and \ref{tab:ngeht-metrics-SGRA} for the M87 and Sgr A* models, respectively. Like \gendirect, we find that the performance of \ehtim\ on M87, once limited to only closure terms and a fixed field-of-view, achieves intermediate performance on image fidelity metrics compared to other submissions. We confirm that the performance of closure-only \ehtim\ on dynamic range shares similar characteristics to the results of \gendirect\ in that they underperform compared to visibility-based reconstructions and there is no significant improvement between the EHT2022 and ngEHT arrays. Nevertheless, \gendirect\ achieves higher relative dynamic range on M87 compared to \ehtim\ on closure terms. Meanwhile for Sgr A* models, \gendirect\ consistently outperforms the reference \ehtim\ output on all datasets for which a successful reconstruction is possible, although the enhancement in quantitative performance is moderate. 

It can be observed that minimising data terms containing closure phases and (log) closure amplitudes does not necessarily help to minimise the closure invariants data term. Alternatively, minimising the closure invariants data term, as we have done with \gendirect, simultaneously minimises data terms with closure phases and amplitudes. By combining closure quantities into the generalised closure invariants formalism, it would be possible to eliminate one degree of freedom in regularised maximum likelihood methods controlling the relative weighting between the phase and amplitude terms. Moreover, when unified as closure invariants, one would not need to apply the usual differential treatment for the uncertainties of closure phases and amplitudes.

\section{Conclusion} \label{sec:conclusion}

Closure invariants are specially constructed interferometric observables composed of combinations of measured visibilities, which capture calibration-independent information about the true source morphology. Recently, \citet{Thyagarajan_2022_CI} and \citet{Samuel_2022} presented a unified formalism of closure invariants for co-polar and polarimetric interferometric measurements, respectively. Later, \citet{Thyagarajan_2024_Lucas} showed that simple machine learning classifiers can use closure invariants to predict an unknown source's morphological class. Then, \citet{Lai_2025_DIReCT} showed that, by leveraging a convolutional autoencoder architecture and the attention mechanism of a trained vision transformer, the complete set of co-polar closure invariants contains sufficient information for generalised direct image reconstruction under the sparse aperture coverage of very-long baseline radio interferometry. However, the \citet{Lai_2025_DIReCT} model performed suboptimally on data metrics because closure terms were not considered as a component of the loss function. Moreover, the model was deterministic, rendering it difficult to interpret the confidence of any individual image reconstruction.

In this work, we presented a generative machine learning approach to interferometric image reconstruction using closure invariants, which can produce a family of images for each provided dataset and minimise the supplied data terms. Below, we summarise the main results:

\begin{itemize}
    \item This work presented \gendirect, a novel machine learning architecture and imaging pipeline designed for VLBI imaging with closure invariants. The model consists of a supervised conditional denoising diffusion UNet and an unsupervised convolutional neural network (CNN). For the predictive pipeline, closure invariants condition the denoising process to sample a set of images, which are used by the CNN to produce the final reconstruction. Once \gendirect\ is trained, there very few hyperparameters that can be tuned for each individual reconstruction. Thus, \gendirect\ provides a relatively simple and easily reproducible imaging tool for very-long baseline interferometry. In following sections, we validated the performance of \gendirect\ on varying levels of thermal noise, as well as on both trained and untrained morphologies.  
    \item Because \gendirect\ operates solely with closure invariants, multiplicative station-based corruptions are canceled out by construction. Therefore, we validate the performance of \gendirect\ on additive thermal noise. Despite being trained on noiseless synthetic observations, the result of \gendirect\ is fairly resilient to high levels of noise, maintaining consistent performance down to a closure invariant $\rm{SNR}\sim3$, implying that \gendirect\ can achieve excellent performance on sources as faint as, and potentially fainter than, M87. 
    \item On all trained and untrained morphologies selected for validation, \gendirect\ produces reconstructions with $\rho_{\rm NX} > 0.9$ on the maximum normalised cross-correlation image metric, for which a value of unity is a perfect reconstruction. On most reconstructions, the performance is $\rho_{\rm NX} \simeq 0.99$, even for a reconstruction of Einstein's face. The data adherence as measured by the $\chi^2_{\rm{CI}}$ metric is also excellent. In the few exceptions, we can recover good performance on $\chi^2_{\rm{CI}}$ and improved $\rho_{\rm NX}$ with minor adjustments. We conclude in this section that as long as much of the overall flux of the observed source is confined within the fixed field-of-view of \gendirect, a high-fidelity reconstruction is possible. 
    \item We leverage the generative diversity of \gendirect\ to quantitatively investigate reconstruction confidence. By aggregating a large number of final reconstructions stemming from a single input dataset, we visualise the median and median absolute deviation images. Furthermore, we define a proxy to visualise the image morphology reconstruction confidence using perceptual hashing methods. We find a high level of consistency throughout the reconstructed sample, with small variations in the image morphology in regions of low surface brightness. 
    \item We apply \gendirect\ on the ngEHT total intensity analysis challenge, consisting of 8 synthetic datasets capturing two horizon-scale source models with two arrays at two different frequencies. We compare \gendirect's performance to closure-only \ehtim\ and other independent challenge submissions, finding that \gendirect\ achieves competitive performance on all datasets as measured by quantitative image and data metrics. Qualitatively, \gendirect\ resolves the ring morphology in seven of eight datasets, with the only exception being insufficiently constrained by the severely sparse aperture coverage. 
\end{itemize}

The performance of \gendirect\ highlights the potential of calibration-independent data terms for co-polar interferometric imaging. Moreover, we demonstrate the utility of novel machine learning methods in solving inverse problems relevant to radio interferometric imaging. In future work, we plan to continuously develop \gendirect\ in the direction of dynamic, polarimetric, and multi-frequency interferometric imaging. Furthermore, the techniques underlying \gendirect\ can be adapted to currently available public EHT data, notably data on M87 which evolve over longer timescales compared to that of the aperture synthesis imaging. The result, which we present in a future study, would constitute an independent reconstruction utilising both novel methods and data terms, offering an additional constraint on source morphology and ultimately enhancing the reliability of sparse VLBI imaging results.

\section*{Acknowledgements}
We thank the anonymous referee for the insightful comments and suggestions, which have improved this manuscript. Inputs from Dominic Pesce and Dongjin Kim are gratefully acknowledged. We also acknowledge Michael Janssen for supplying the FITS model image of Centaurus A used in this study and Freek Roelofs for providing access to the ngEHT Challenge datasets. 

Software packages used in this study include \textsc{Numpy} \citep{Numpy_2011}, \textsc{Scipy} \citep{Scipy_2020}, \textsc{PyTorch} \citep{PyTorch}, \ehtim\ \citep{Chael_2018_ehtim}, \gendirect\ \citep{GenDIReCT_Zenodo}, \textsc{ClosureInvariants} \citep{CI_Package}, and \textsc{Matplotlib} \citep{Matplotlib_2007}.

\paragraph{Funding Statement}
None

\paragraph{Competing Interests}
None

\paragraph{Data Availability Statement}
The data underlying this article will be shared on reasonable request to the corresponding author. Code for \gendirect\ is publicly accessible on GitHub\footnote{\href{https://github.com/samlaihei/GenDIReCT}{https://github.com/samlaihei/GenDIReCT}}.

\paragraph{Ethical Standards}
The research meets all ethical guidelines, including adherence to the legal requirements of the study country.

\paragraph{Author Contributions}
Conceptualization: S.L; N.T. Methodology: S.L; F.D; N.T; I.W. Data curation: S.L. Data visualisation: S.L. Writing original draft: S.L. All authors approved the final submitted draft.

\printendnotes

\bibliography{bibliography}
\appendix

\section{The diffusion objective} \label{appendix:diffobj}
We first begin with a brief summary of the principles behind denoising diffusion probabilistic models as described in the seminal paper, \citet{DDPM_ho_2020}. Suppose that an image $x_0$ is sampled from an underlying distribution $q(x_0)$. Forward diffusion on an image is a Markov chain process which advances from timestep $t=0$ to $T$, defined as,
\begin{equation}
    q(x_{t}|x_{t-1}) = \mathcal{N}(x_{t};\mu_t,\sigma^2_t)\,,
\end{equation}
where $\mu_t = \sqrt{1-\beta_t}\, x_{t-1}$ and $\sigma_t^2 = \beta_t \mathcal{I}$ for a time-dependent variance $\beta_t$. A normal distribution is denoted by $\mathcal{N}$ and $\mathcal{I}$ is the identity matrix. After $T\gg 1$ timesteps, $x_T$ becomes consistent with a pure isotropic Gaussian distribution. Therefore, given $x_{t-1}$, one could produce the new noisier image under forward diffusion with $x_t = \sqrt{1-\beta_t}x_{t-1} + \sqrt{\beta_t}\epsilon_t$, with $\epsilon_t \sim \mathcal{N}(0, \mathcal{I})$. By propagating the Markov chain forward from $x_0$, it can be shown that the relationship between $x_t$ to $x_0$ can be expressed directly through reparameterisation as $x_t = \sqrt{\bar{a_t}}x_0 + \sqrt{1-\bar{a_t}}\epsilon_t$, where $a_t = 1-\beta_t$ and $\bar{a_t} = \Pi_{i=1}^{t}a_i$ \citep{Sohl-Dickstein_2015}. Because $\bar{a_t}$ is only dependent on the predetermined variance schedule $\beta_t$, it can be precomputed to enable estimating $x_t$ from $x_0$ in a single sampling of an appropriately scaled Gaussian noise distribution.

The objective of a diffusion model is to approximate the conditional probability distribution $p(x_{t-1}|x_{t})$, representing the reverse diffusive process, using a deep neural network. Under the assumption that the reverse process is also Gaussian, we can approximate the conditional probability distribution with a learnable mean $\mu_\theta$ and variance $\sigma^2_\theta$, 
\begin{equation}
    p_{\theta}(x_{t-1}|x_{t}) = \mathcal{N}\left(x_{t-1};\mu_\theta(x_t, t),\sigma^2_\theta(x_t, t)\right)\,,
\end{equation}
where $\theta$ describes the parameters of the neural network optimised by gradient descent. \citet{Luo_2022_UnderstandingDiffusion} showed that the reparameterised mean can be expressed as,
\begin{equation}
    \mu_\theta(x_t, t) = \frac{1}{\sqrt{a_t}}\left(x_t - \frac{\beta_t}{\sqrt{1-\bar{a_t}}}\epsilon_\theta(x_t,t)\right)\,,
\end{equation}
where $\epsilon_\theta(x_t,t)$ is the Gaussian noise predicted by the network for a particular dataset $x_t$ and timestep $t$. Therefore, the loss between $\mu_\theta$ and $\mu_t$ can be expressed solely in terms of the applied noise $\epsilon_t$ and predicted noise $\epsilon_\theta(x_t, t)$. Although $\sigma^2_\theta(x_t,t)$ is also in principle learnable by the network \citep{Nichol_2021_ImprovedDDPM}, we follow after \citet{DDPM_ho_2020} and \citet{Rombach_2021_StableDiff} in optimising a simplified form of the objective function $\mathcal{L}_{\rm{simple}}$, which achieves comparable results or even outperforms the full objective function,
\begin{equation}
    \begin{aligned}
    \mathcal{L}_{\rm{simple}} &= \mathbbm{E}_{x_t, t, \epsilon_t}\left[\left\Vert \epsilon_t - \epsilon_\theta\left(x_t, t\right)\right\Vert^2\right]\,,\\
     &= \mathbbm{E}_{x_0, t, \epsilon_t}\left[\left\Vert \epsilon_t - \epsilon_\theta\left(\sqrt{\bar{a_t}}x_0 + \sqrt{1-\bar{a_t}} \epsilon_t, t\right)\right\Vert^2\right]\,,
    \end{aligned}
\end{equation}
where $\mathbbm{E}$ is the expectation operator. By taking gradient descent steps on the above objective function, as $\nabla_\theta\mathcal{L}_{\rm{simple}}$, we can train a neural network model to generate images that exhibit many qualities similar to those in the training dataset using reverse diffusion from a purely isotropic noise input. This framework has also been explored by \citet{Feng_2024} to generate data-driven image priors for a variational Bayesian imaging method utilising normalising flow techniques. 

In order to generate samples consistent with a particular input, we need to manipulate the reverse diffusion process by conditioning the prior data distribution with additional parameters, $y$, as $p_{\theta}(x_{t-1}|x_{t}, y)$. We follow after \citet{Rombach_2021_StableDiff} by mapping the conditioning features to intermediate layers of the UNet via cross-attention layers \citep{Vaswani_2017, Devlin_2018_BERT}, thereby optimising the conditional objective function,
\begin{equation}
    \mathcal{L}_{\rm{\gendirect}} = \mathbbm{E}_{x_t, y, t, \epsilon_t}\left[\left\Vert \epsilon_t - \epsilon_\theta\left(x_t, t, y\right)\right\Vert^2\right], 
\end{equation}
which ensures that the learned conditional image distribution, $p_\theta(x|y)$, approximates the training data conditional distribution, $p_{\rm data}(x|y)$. We note that \citet{Rombach_2021_StableDiff} also describes a domain-specific pre-trained encoder, $\tau_\theta(y)$, which is optimised simultaneously alongside the diffusion model, $\epsilon_\theta$. However, we find that conditioning the network on the $y$ features directly results in better performance in our evaluations.

Regarding the architecture of the diffusion model, we employ the UNet model \citep{Ronneberger_2015_Unet}, which consists of convolutional layers set in a symmetric contracting and expanding pathways with residual connections \citep{He_2016_RESNET} from the contracting path that map features to their mirrored expanding layers. The timestep, $t$, is encoded with the input by employing sinusoidal positional embeddings \citep[e.g.][]{Vaswani_2017}. We present a visual representation of the employed UNet architecture in Figure \ref{fig:gendirect-architecture}, which illustrates how the forward diffusion process is applied to the encoded latent features of input images from the training dataset. Each diffusion step is predicted by the conditional denoising UNet using the objective function of Equation~(\ref{eq:gendirect-loss}). Within the conditional denoising UNet, features from closure invariants are concatenated with some of the convolutional layers via the cross-attention mechanism, which conditions the denoising process based on the input closure invariants. Additionally, though it is not explicitly illustrated, the UNet is designed with self-attention layers and standard convolutional layers without attention.


\section{Data compression by averaging closure invariants} \label{appendix:avg-ci}
Throughout Sections \ref{sec:results} and \ref{sec:discussion}, we have compressed the visibility information on 10-minute scans, where within each scan, data is captured on a 10s cadence. By aggregating information in this way, we can compress the dataset into a more manageable volume and improve the SNR. However, visibilities can be influenced by more sources of data corruption than closure invariants. In this section, we show that averaging closure invariants is generally a more robust method of aggregating data than averaging visibilities prior to computing closure invariants in the presence of noise. We also discuss the circumstances for which the difference in the outcome between the two methods can be substantial. 

To demonstrate the difference between the two methods, we run a simulation where we measure the sum of squared errors (SSE) between noisy and true closure invariants resulting from both processes. We utilise \ehtim\ to synthetically perform a 12-hour rotation synthesis noisy observation using stations from the EHT array with standard SEFDs, capturing data at a 100s cadence, and using the same Sgr A* model as in Figure \ref{fig:gendirect-outputs} and Section \ref{sec:thermal-noise}. In the first case, we average the visibilities across 600s windows before computing the closure invariants and in the second case, we compute all of the closure invariants from the pre-processed visibilities before averaging them over 600s windows. 

\begin{figure}
	\includegraphics[width=0.9\textwidth]{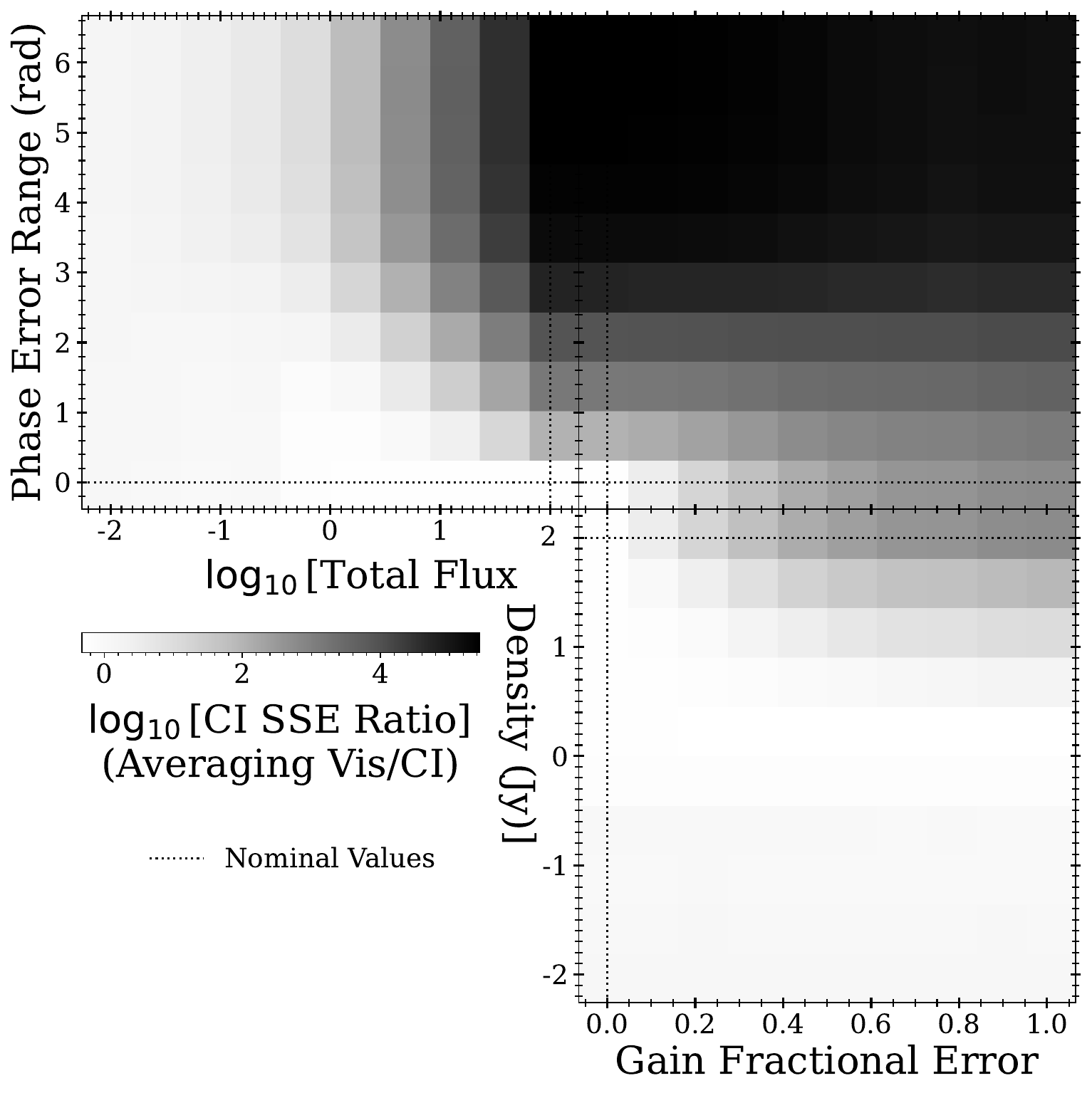}
    \caption[]{A multi-dimensional covariant illustration of the SSE ratio between averaging visibilities and closure invariants over the total flux density of the source, time-dependent multiplicative gain corruption, and phase corruption. Each panel presents the variation in SSE ratio over two parameters, keeping the third parameter fixed at the nominal values of 100 Jy, 0 gain error, or 0 phase error, as indicated by the dotted lines. }
    \label{fig:averaging-ci}
\end{figure}

\begin{figure*}
	\includegraphics[width=0.9\textwidth]{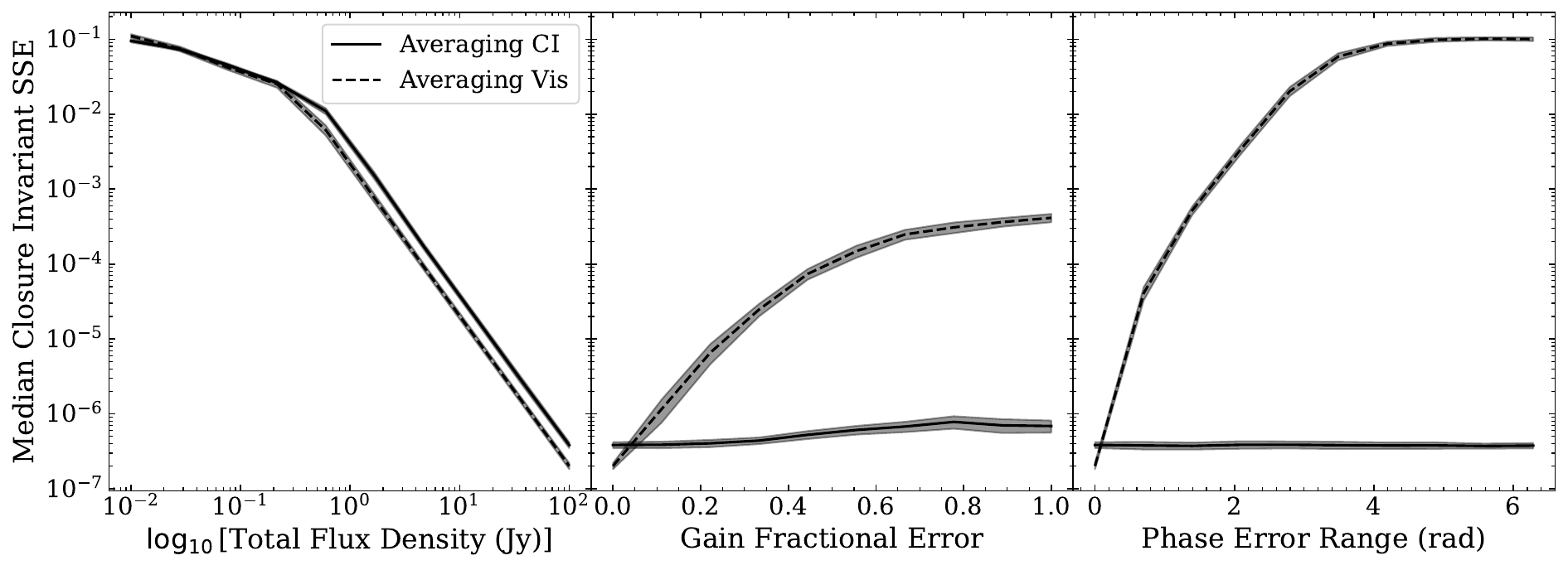}
    \caption[]{Relationship between the median closure invariant SSE with the total flux density (left), time-dependent gain error (middle), and phase error (left). In each panel, the other two parameters are fixed at nominal values of 100 Jy, 0 gain error, and 0 phase error to help isolate the effect of varying each parameter independently. The solid and dashed lines correspond to the closure invariants SSE obtained from closure averaging and visibility averaging, respectively.}
    \label{fig:averaging-ci-1d}
\end{figure*}

Figure \ref{fig:averaging-ci} presents the median SSE ratio between the two methods by varying the flux density of the source (a proxy for signal-to-noise ratio for a fixed set of SEFDs), time-dependent multiplicative amplitude corruption, and phase corruption. In each panel illustrating the covariance between two parameters, the third parameter is fixed at nominal values of 100 Jy (a proxy for highest signal-to-noise ratio), $0$ amplitude error, or $0^\circ$ phase error. Likewise, in Figure \ref{fig:averaging-ci-1d}, two of the parameters remain fixed at nominal values while the effect of varying one parameter on the median closure invariant SSE is illustrated independently for both strategies. The advantage conferred by averaging closure invariants over visibilities is most sensitive to multiplicative corruptions, while the alternative strategy of averaging visibilities would only be preferable for a bright source with perfectly calibrated visibilities. Thus, whenever data compression is necessary in future applications involving closure invariants, aggregating the closure quantities rather than visibilities generally results in more robust data terms. Adopting this strategy has the potential to further improve the resilience of \gendirect\ to realistic noise corruption (additive and multiplicative). However, in this work, where we have assumed no time-dependent multiplicative corruptions and principally explored the consequences of enhanced thermal noise corruption on our reconstruction fidelity for 1 Jy sources in Section \ref{sec:thermal-noise}, obtaining closure invariants after aggregating visibilities would not present a significant disadvantage nor affect the quality of image reconstructions in this work. We further note that if the multiplicative corruptions are stable across the averaging timescale, as is the case for the simulated ngEHT total intensity analysis challenge data, then the disadvantage conferred by obtaining closure invariants after averaging visibilities is insignificant. 

\end{document}